\documentclass[useAMS,usenatbib,usegraphicx]{mn2e}

\usepackage[T1]{fontenc}
\usepackage{pslatex}
\usepackage{color}
\usepackage{amsmath}
\usepackage{url}

\def\gtrsim{\mathrel{\hbox{\rlap{\hbox{\lower3pt\hbox{$\sim$}}}\hbox{\raise2pt\hbox{$>$}}}}}

\newcommand{\nodata}{...}

\title[Far-IR properties of $z\sim2$ GNS galaxies]{\textit{Herschel} observations of a $\bf{z\sim 2}$ stellar mass 
selected galaxy sample drawn from the GOODS NICMOS Survey}

\author[M.~Hilton et al.]
{\parbox{\textwidth}{\raggedright M.~Hilton,$^{1}$\thanks{E-mail: matthew.hilton@nottingham.ac.uk}
C.J.~Conselice,$^{1}$
I.G.~Roseboom,$^{2,3}$
D.~Burgarella,$^{4}$
V.~Buat,$^{4}$
S.~Berta,$^{5}$
M.~B{\'e}thermin,$^{6,7}$
J.~Bock,$^{8,9}$
S.C.~Chapman,$^{10}$
D.L.~Clements,$^{11}$
A.~Conley,$^{12}$
L.~Conversi,$^{13}$
A.~Cooray,$^{14,8}$
D.~Farrah,$^{2,15}$
E.~Ibar,$^{16}$
G.~Magdis,$^{17}$
B.~Magnelli,$^{5}$
G.~Marsden,$^{18}$
R.~Nordon,$^{5}$
S.J.~Oliver,$^{2}$
M.J.~Page,$^{19}$
P.~Popesso,$^{5}$
F.~Pozzi,$^{20}$
B.~Schulz,$^{8,21}$
Douglas~Scott,$^{18}$
A.J.~Smith,$^{2}$
M.~Symeonidis,$^{19}$
I.~Valtchanov,$^{13}$
M.~Viero,$^{8}$
L.~Wang$^{2}$ and
M.~Zemcov$^{8,9}$}\vspace{0.4cm}\\
\parbox{\textwidth}{\raggedright $^{1}$Centre for Astronomy \& Particle Theory, School of Physics and Astronomy, University of Nottingham, NG7 2RD, UK\\
$^{2}$Astronomy Centre, Dept. of Physics \& Astronomy, University of Sussex, Brighton BN1 9QH, UK\\
$^{3}$Institute for Astronomy, University of Edinburgh, Royal Observatory, Blackford Hill, Edinburgh EH9 3HJ, UK\\
$^{4}$Laboratoire d'Astrophysique de Marseille, OAMP, Universit\'e Aix-marseille, CNRS, 38 rue Fr\'ed\'eric Joliot-Curie, 13388 Marseille cedex 13, France\\
$^{5}$Max-Planck-Institut f\"ur Extraterrestrische Physik (MPE), Postfach 1312, 85741, Garching, Germany\\
$^{6}$Laboratoire AIM-Paris-Saclay, CEA/DSM/Irfu - CNRS - Universit\'e Paris Diderot, CE-Saclay, pt courrier 131, F-91191 Gif-sur-Yvette, France\\
$^{7}$Institut d'Astrophysique Spatiale (IAS), b\^atiment 121, Universit\'e Paris-Sud 11 and CNRS (UMR 8617), 91405 Orsay, France\\
$^{8}$California Institute of Technology, 1200 E. California Blvd., Pasadena, CA 91125, USA\\
$^{9}$Jet Propulsion Laboratory, 4800 Oak Grove Drive, Pasadena, CA 91109, USA\\
$^{10}$Institute of Astronomy, University of Cambridge, Madingley Road, Cambridge CB3 0HA, UK\\
$^{11}$Astrophysics Group, Imperial College London, Blackett Laboratory, Prince Consort Road, London SW7 2AZ, UK\\
$^{12}$Center for Astrophysics and Space Astronomy 389-UCB, University of Colorado, Boulder, CO 80309, USA\\
$^{13}$Herschel Science Centre, European Space Astronomy Centre, Villanueva de la Ca\~nada, 28691 Madrid, Spain\\
$^{14}$Dept. of Physics \& Astronomy, University of California, Irvine, CA 92697, USA\\
$^{15}$Department of Physics, Virginia Tech, Blacksburg, VA 24061, USA\\
$^{16}$UK Astronomy Technology Centre, Royal Observatory, Blackford Hill, Edinburgh EH9 3HJ, UK\\
$^{17}$Department of Astrophysics, Denys Wilkinson Building, University of Oxford, Keble Road, Oxford OX1 3RH, UK\\
$^{18}$Department of Physics \& Astronomy, University of British Columbia, 6224 Agricultural Road, Vancouver, BC V6T~1Z1, Canada\\
$^{19}$Mullard Space Science Laboratory, University College London, Holmbury St. Mary, Dorking, Surrey RH5 6NT, UK\\
$^{20}$INAF-Osservatorio Astronomico di Roma, via di Franscati 33, 00040 Monte Porzio Catone, Italy\\
$^{21}$Infrared Processing and Analysis Center, MS 100-22, California Institute of Technology, JPL, Pasadena, CA 91125, USA}}

\begin{document}

\date{Draft version: \today}

\pagerange{\pageref{firstpage}--\pageref{lastpage}} \pubyear{2012}

\maketitle

\label{firstpage}

\begin{abstract}
We present a study of the far-IR properties of a stellar mass selected sample of $1.5 < z < 3$
galaxies with $\log (M_*/{\rm M_{\sun}}) > 9.5$ drawn from the GOODS NICMOS Survey (GNS), the deepest 
$H$-band \textit{Hubble Space Telescope} survey of its type prior to the installation of WFC3. We use far-IR and
sub-mm data from the PACS and SPIRE instruments on-board \textit{Herschel}, taken from 
the PACS Evolutionary Probe (PEP) and \textit{Herschel} Multi-Tiered Extragalactic Survey (HerMES) key projects
respectively. We find a total of 22 GNS galaxies, with median $\log (M_*/{\rm M_{\sun}}) = 10.8$ and
$z = 2.0$, associated with 250$~\micron$ sources detected with SNR $> 3$. We derive mean total IR luminosity $\log L_{\rm IR} (\rm L_{\sun}) = 12.36
\pm 0.05$ and corresponding star formation rate SFR$_{\rm IR+UV}$ $=(280 \pm 40)$\,$\rm M_{\sun}$~yr$^{-1}$ for
these objects, and find them to have mean dust temperature $T_{\rm dust} \approx 35$\,K. We find that the SFR
derived from the far-IR photometry combined with UV-based estimates of 
unobscured SFR for these galaxies is on average more than a factor of 2 higher than the SFR derived from 
extinction corrected UV emission alone, although we note that the IR-based estimate is subject to substantial
Malmquist bias. To mitigate the effect of this bias and extend our study to fainter fluxes, we perform a stacking 
analysis to measure the mean SFR in bins of stellar mass. We obtain
detections at the $2-4\sigma$ level at SPIRE wavelengths for samples with $\log (M_*/{\rm M_{\sun}}) > 10$. In
contrast to the \textit{Herschel} detected GNS galaxies, we find that estimates of SFR$_{\rm IR+UV}$ for the
stacked samples are comparable to those derived from extinction corrected UV emission, although
the uncertainties are large. We find evidence for an increasing fraction of dust obscured star formation
with stellar mass, finding SFR$_{\rm IR}/$SFR$_{\rm UV} \propto M_*^{0.7 \pm 0.2}$, which is likely a consequence
of the mass--metallicity relation.
\end{abstract}
\begin{keywords}
galaxies: evolution -- galaxies: high-redshift -- galaxies: starburst -- infrared: galaxies
\end{keywords}

\section{Introduction}
\label{s_intro}
Star formation rates (SFRs) in galaxies can be measured using many different methods \citep[see
e.g.][]{Kennicutt_1998}. The most easily accessible tracer at high-redshift ($z > 1$) is rest-frame UV
emission, which correlates with the number of young, massive stars and hence the global SFR of
a galaxy. However, in dusty galaxies, this requires a significant correction due to absorption of UV
photons by dust, which can be estimated using the correlation between the UV and far-infrared (IR) luminosity
ratio ($L_{\rm IR}/L_{\rm UV}$, where $L_{\rm IR}$ is conventionally defined over the wavelength range $8-1000$\,$\micron$)
and the UV slope ($\beta$; typically determined from a power law fit of the form $f_\lambda \propto \lambda^{\beta}$
between 1500 and 2800\,\AA{}), which has been measured from local
starburst galaxies \citep[e.g.][]{Meurer_1999, Calzetti_2000}. Observations
at far-IR wavelengths are generally thought to quantify the amount of obscured star formation more directly, 
as UV radiation associated with young stellar populations is absorbed by interstellar dust and re-emitted at 
far-IR wavelengths, and have revealed that much of the star formation activity that occurred at $z > 1$ is 
obscured \citep[e.g.][]{PerezGonzalez_2005, LeFloc_2005, Caputi_2007, Magnelli_2009, Magnelli_2011}.

Observations over the last decade spanning a wide range in redshift and galaxy environments have
shown that stellar mass is a key parameter for predicting the properties of a given galaxy. At low redshift
($z < 0.1$), the most massive galaxies tend to be red and located in denser environments than bluer, lower
mass galaxies \citep[e.g.][]{Baldry_2006}. Although the colour--density relation weakens as redshift
increases, a strong colour--mass relation is still seen at $z \sim 2$ \citep[e.g.][]{Gruetzbauch_2011}. For
galaxies which are actively forming stars, SFR is seen to be correlated with stellar mass up to $z \sim 3$
\citep[e.g.][]{Daddi_2007I, Magdis_2010, Oliver_2010, Bauer_2011, Karim_2011, Rodighiero_2011}. Environment,
while
certainly important (as seen by the dominance of early type, passively evolving galaxies in clusters), seems
to be more weakly correlated with other galaxy properties in comparison to stellar mass, particularly at high
redshift \citep[e.g.][]{Peng_2010, GruetzbauchSFR_2011}. This suggests that studies of the assembly of stellar
mass, much of which occurs in obscured bursts of star formation, are crucial for developing our understanding
of the galaxy formation process.

In this paper we use far-IR photometry from the \textit{Herschel Space Observatory} \citep{Pilbratt_2010} HerMES \citep{Oliver_2011}
and PEP \citep{Lutz_2011} key projects to investigate obscured star formation in a stellar mass selected galaxy sample: the GOODS NICMOS Survey
\citep[GNS;][]{Conselice_2011}. The GNS sample is selected in the $H$-band and is estimated to be complete for
galaxies with stellar masses down to $\log (M_*/{\rm M_{\sun}}) = 9.5$ at $z < 3$ \citep[][]{Gruetzbauch_2011,
Mortlock_2011, Conselice_2011}. \citet{Bauer_2011} carried out a
study of star formation activity in the GNS sample over the redshift range $1.5 < z < 3$. This coincides with
the peak of cosmic star formation activity as measured in the UV \citep[e.g.][]{Bouwens_2009}; note however
that in the IR a flat plateau in the SFR density is seen from $1 < z < 2$ \citep[e.g.][]{Magnelli_2011, 
Bethermin_2011}. The \citet{Bauer_2011} study primarily used rest-frame UV luminosity (corrected
for extinction according to the UV slope) to estimate SFRs. In addition, they estimated
obscured SFRs for the $\approx 20$ per cent of their sample that were detected at 24\,$\micron$ using the MIPS
instrument on board \textit{Spitzer}, finding that the inferred total star formation rate (SFR$_{\rm IR+UV}$)
is on average 3.5 times larger than the SFR derived from the UV-slope extinction corrected
UV flux (SFR$_{\rm UV,corr}$). This factor of 3.5 may be overestimated, as several previous studies have shown that
while 24\,$\micron$ flux densities can be reasonably extrapolated to measure $L_{\rm IR}$ (and hence SFR$_{\rm IR}$)
for galaxies at $z < 1.5$, this is not the case at higher redshift \citep[e.g.][]{Papovich_2007, Murphy_2009,
Murphy_2011}, where $L_{\rm IR}$ as estimated from 24\,$\micron$ photometry alone can be a factor $\sim 5$
higher than $L_{\rm IR}$ measured for the same sources when additional longer wavelength photometry is 
available to constrain the SED fits. The discrepancy is greater for Ultra Luminous Infrared
Galaxies (ULIRGs, which have $L_{\rm IR} > 10^{12}$~$\rm L_{\sun}$). Similar results have been reported in studies
using \textit{Herschel} data \citep[e.g.][]{Elbaz_2010, Elbaz_2011, Nordon_2010, Nordon_2011}.

\begin{figure*}
\includegraphics[height=8.5cm]{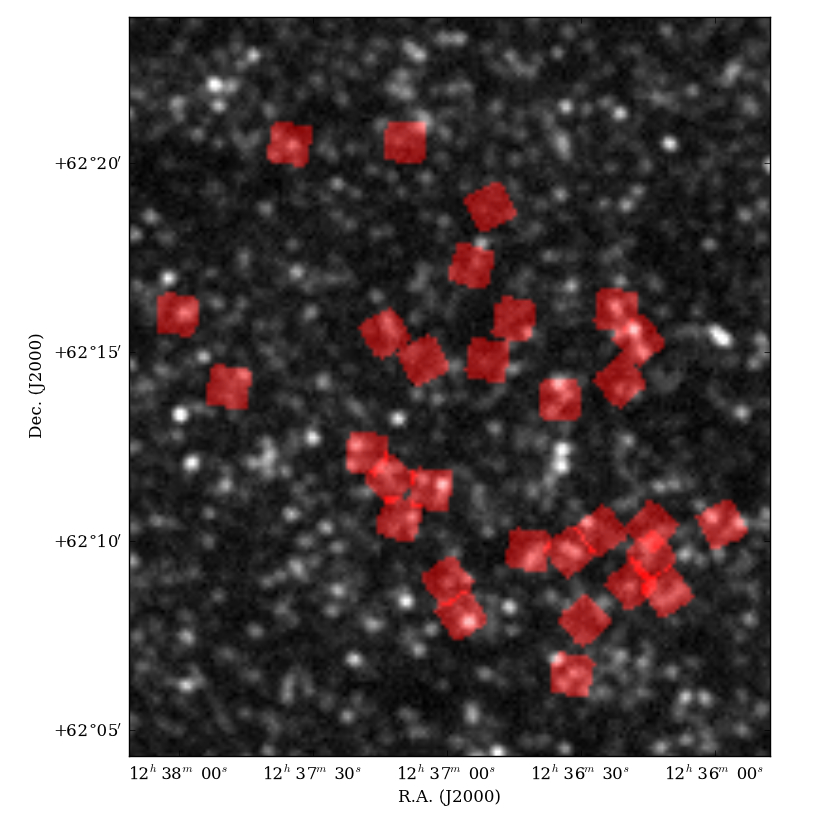}
\includegraphics[height=8.5cm]{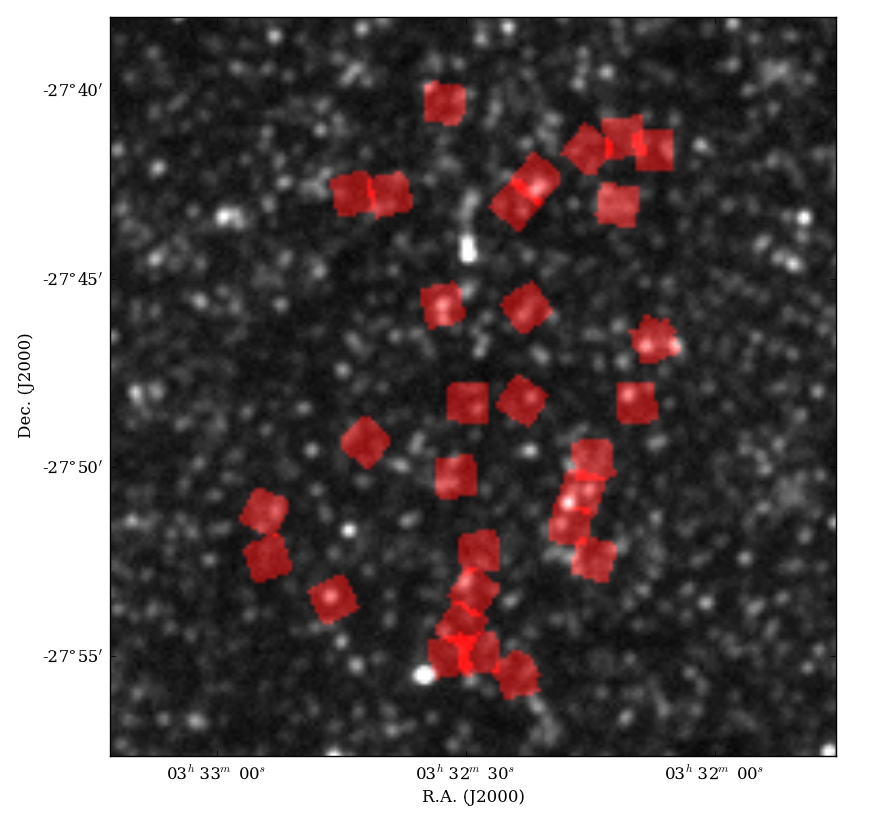}
\caption{Footprint of the GOODS NICMOS Survey (red) overlaid on the HerMES 250\,$\micron$ maps of the
GOODS-North (left) and GOODS-South (right) fields. Each GNS pointing is in the direction of one or more $M_* >
10^{11}$\,M$_{\sun}$ galaxies at $1.7 < z < 2.9$, and is about $50\arcsec$ on a side.}
\label{f_GNSOverlay}
\end{figure*}

Star formation in the massive ($M_* > 10^{11}$~$\rm M_{\sun}$) galaxies on which most of the GNS fields are
centred (see Section~\ref{s_sample}) has been investigated using
far-IR data from the Balloon-borne Large-Aperture Submillimeter Telescope \citep[BLAST;][]{Viero_2010} and
\textit{Herschel} \citep{Cava_2010}, who found that disk-like galaxies (selected by the use of the S\'{e}rsic
index) have significantly higher SFRs than spheroidal-like galaxies. In this work we aim to improve the
characterisation of obscured star formation as a function of stellar mass at $1.5 < z < 3$, using the combination of
\textit{Herschel} photometry and the wide stellar mass range spanned by the full GNS sample ($\log (M_*/{\rm
M_{\sun}}) > 9.5$).

The structure of this paper is as follows. In Section~\ref{s_data} we give a brief overview of the GNS and the
\textit{Herschel} data used in this work. We investigate the properties of the GNS galaxies detected
at 250~$\mu$m using \textit{Herschel} in Section~\ref{s_detections}. We extend the study to lower luminosity
galaxies through a stacking analysis which is presented in Section~\ref{s_stacking}. We present our
conclusions in Section~\ref{s_conclusions}.

We assume a cosmology with $\Omega_{\rm m}=0.3$, $\Omega_\Lambda=0.7$, and $H_0=70$ km s$^{-1}$ Mpc$^{-1}$
throughout. All values for star formation rates and stellar masses assume a \citet{Chabrier_2003} Initial Mass
Function (IMF), unless noted otherwise.

\section{Data}
\label{s_data}

\subsection{Galaxy sample}
\label{s_sample}
The galaxy sample used in this work is taken from the GNS \citep{Conselice_2011}, which consists of 60 F160W
($H$-band) pointed observations in the GOODS fields \citep{Giavalisco_2004} using the NICMOS instrument
on-board the \textit{Hubble Space Telescope} (\textit{HST}). The footprint of the GNS overlaid on the SPIRE
250\,$\micron$ maps is shown in Fig.~\ref{f_GNSOverlay}. Each GNS field is $\approx 50\arcsec$ on a side,
and covers the region around one or more massive galaxies ($M_* > 10^{11}$\,M$_{\sun}$) at $1.7 < z < 2.9$,
initially selected using a variety of colour selection techniques: distant red galaxies
\citep[DRGs;][]{Papovich_2006}, IRAC extremely red objects \citep[IEROs;][]{Yan_2004} and BzK galaxies
\citep{Daddi_2007I}. While this selection is not homogeneous, \citet{Conselice_2011} shows that this
combination of colour selection techniques leads to an almost complete sample of massive ($M_* >
10^{11}$\,M$_{\sun}$) galaxies: no single one of these colour selection methods selects more than 70 per cent of the
massive galaxy population that would be selected in a photometric redshift survey, while a
subsequent stellar mass selection in these fields based on photometric redshifts 
found an almost identical massive galaxy sample to the initial colour-based selection \citep{Conselice_2011}.

In addition to providing high-resolution near-IR photometry of the massive galaxies targeted in each GNS
pointing, the depth of the survey allows galaxies with much lower stellar masses to be detected: GNS is
complete for galaxies with stellar masses down to $\log (M_*/{\rm M_{\sun}}) = 9.5$ at $z < 3$
\citep[][]{Gruetzbauch_2011, Mortlock_2011}. The stellar mass measurements are described in detail in
\citet{Conselice_2011}; briefly, a grid of \citet{BruzualCharlot_2003} stellar population models, with
exponentially declining star formation histories ($\tau$-models, with $0.01 < \tau (\rm Gyr) < 10$), spanning
a wide range in metallicity ($-2.25 < \rm [Fe/H] < +0.56$), were fitted to the $BVizH$ photometry for each
galaxy. 

In this paper we use a sample of 860 $1.5 < z < 3$ galaxies with $\log (M_*/{\rm M_{\sun}}) > 9.5$ drawn from the GNS 
(the redshift range is chosen to match previous analyses of this catalogue presented in, e.g. 
\citealt{Bauer_2011, GruetzbauchSFR_2011, Mortlock_2011}). We include galaxies with both spectroscopic and 
photometric redshifts, using the former where possible. We do not cut galaxies with low photometric redshift 
probability ($P$, the $\chi^2$ probability outputted by \texttt{HYPERZ}, the code used to 
compute the GNS photometric redshifts; \citealt{Bolzonella_2000}),
because a comparison of the spectroscopic and photometric redshifts showed that the scatter of the residuals
is similar regardless of the cut in $P$ ($\sigma_z = 0.045$\footnotemark when using only galaxies with $P > 95$ per cent,
compared to $\sigma_z = 0.06$ using the full sample; see \citealt{Gruetzbauch_2011, Bauer_2011}). Note that
450 galaxies in this sample have $P > 95$ per cent.
\footnotetext{$\sigma_z$ is defined as the scatter in the photometric redshift residuals, i.e. 
$\delta z = (z_{\rm spec} - z_{\rm phot}) / (1 + z_{\rm spec})$.}

To reduce contamination of the sample by AGN, we remove galaxies found within a 2$\arcsec$ matching radius of
X-ray sources listed in the 2~Msec \textit{Chandra} catalogues of \citet[][GOODS-North]{Alexander_2003} and
\citet[][GOODS-South]{Luo_2008}. These catalogues have flux limits of $\approx 1.4 \times
10^{-16}$~ergs~cm$^{-2}$~s$^{-1}$ in the 2-8\,keV band, and are therefore deep enough to allow sources
brighter than $L_{\rm X(2-8\, keV)} \sim 4 \times 10^{42}$~ergs~s$^{-1}$ to be detected at $z \sim 2$ 
(assuming a power law spectrum with $\Gamma = 2$).

\begin{figure*}
\includegraphics[width=18cm]{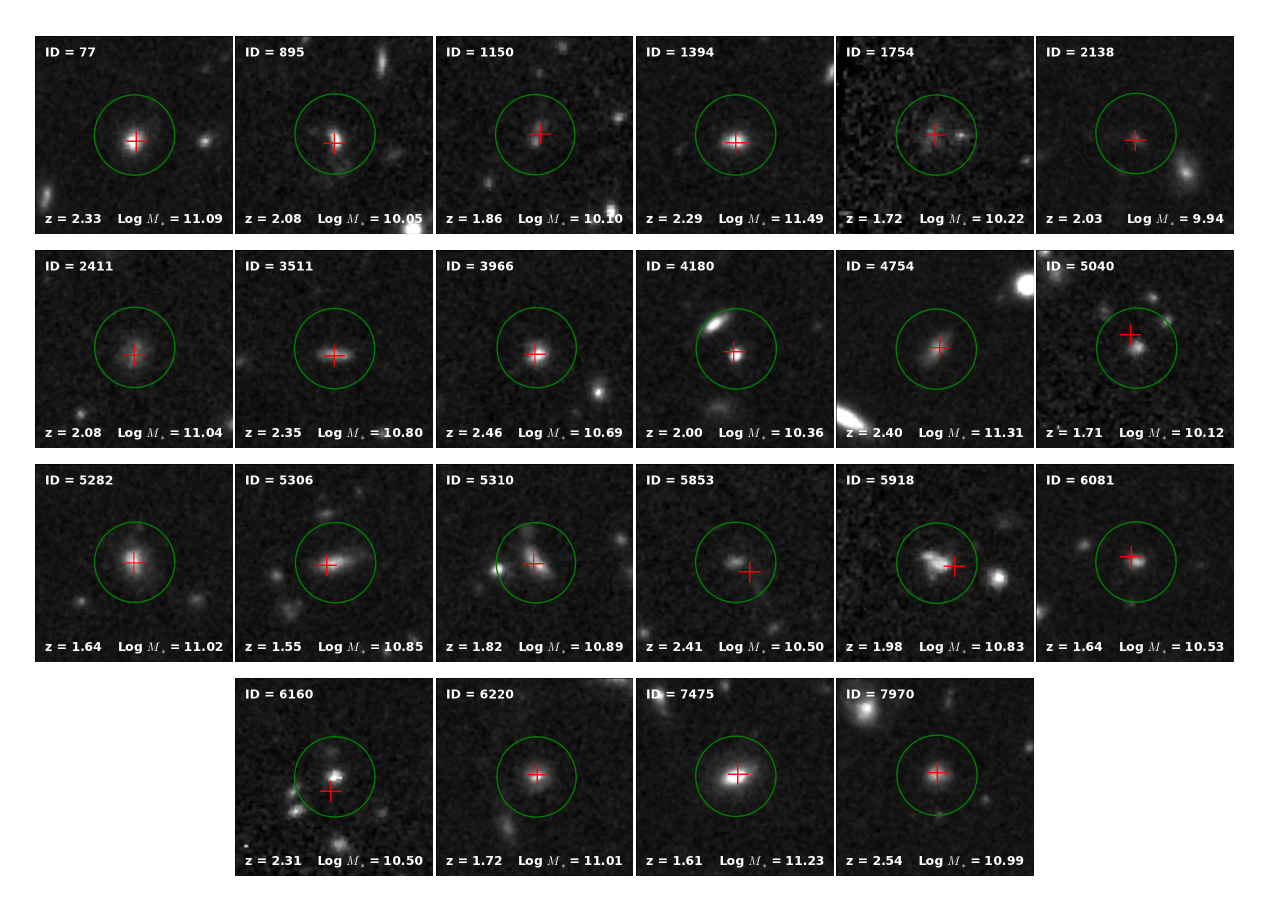}
\caption{Postage stamp ($10\arcsec \times 10\arcsec$) NICMOS F160W ($H$-band) images of GNS galaxies detected
in HerMES with SNR $> 3$ at 250\,$\micron$. The red cross in each postage stamp marks the position of the
corresponding matched object in the HerMES/PEP catalogue, which is extracted using MIPS 24\,$\micron$
priors. The green circle indicates the 2$\arcsec$ matching radius used for cross-matching between the two
catalogues.}
\label{f_postageStamps}
\end{figure*}

Later in this paper, we measure SFRs for GNS galaxies from the \textit{Herschel} IR data and compare these
with UV-based SFR measurements from \citet{Bauer_2011} for the same galaxy sample. Here, we briefly
summarise the method used to estimate these UV-based SFRs.

\citet{Bauer_2011} estimated unobscured UV SFRs from $K$-corrected ACS
$z_{850}$-band flux measurements, applying the SFR$_{\rm UV}$-$L_{2800}$ relation of \citet{Kennicutt_1998},
where $L_{2800}$ is the UV luminosity at 2800\,\AA{}. These were corrected for obscuration by dust using the UV
slope ($\beta$) to estimate the amount of extinction, where $\beta$ was measured from the 1600\,\AA{} and
2800\,\AA{} luminosities of the best fitting model SED for each galaxy. A similar methodology to
\citet{Calzetti_2000} was used to convert $\beta$ values into extinction estimates at 2800\,\AA{}. The typical
uncertainty on the UV-slope extinction corrected SFR estimates (SFR$_{\rm UV,corr}$) is $\sim 30$ per cent
\citep{Bauer_2011}.

\subsection{Infrared data}
\label{s_IRData}
The \textit{Herschel} photometry used in this work is taken from two key projects. The PACS \citep{Poglitsch_2010}
Evolutionary
Probe \citep[PEP;][]{Lutz_2011} provides 100 and 160\,$\micron$ data covering both GOODS fields, as well as
70\,$\micron$ coverage of GOODS-S. Simulations show that in GOODS-N, the flux limits at 80 per cent
completeness are 4.5 and 7.0\,mJy at 100 and 160\,$\micron$ respectively, while in GOODS-S the corresponding
limits are 1.5, 2.0, and 4.8\,mJy at 70, 100, and 160\,$\micron$. We also use 250, 350, and 500\,$\micron$ SPIRE
\citep{Griffin_2010} imaging data which was obtained as part of the \textit{Herschel} Multi-Tiered
Extragalactic Survey \citep[HerMES\footnotemark;][]{Oliver_2010b, Oliver_2011}. Unlike the PACS data, the SPIRE data are
dominated by confusion noise from unresolved background sources. The calibration of the SPIRE instrument is
described in \citet{Swinyard_2010}. 
\footnotetext{\url{http://hermes.sussex.ac.uk}}

Photometry was performed on all the \textit{Herschel} maps, using prior positions derived from the MIPS
24\,$\micron$ catalogue of \citet{Magnelli_2009} for source extraction. This 24\,$\micron$ catalogue is 
extracted from the GOODS-Legacy program observations (PI: M. Dickinson), and reaches a 5$\sigma$ depth of about 
30\,$\mu$Jy. Note that by requiring a
24\,$\micron$ detection for source extraction in the \textit{Herschel} maps, a small fraction
of sources will be missed at the GOODS depth \citep[$< 10$ per cent; e.g.][]{Roseboom_2010, 
Magdis_2011, Bethermin_2012}. A blind extraction might be able to
find such sources, at the expense of significantly noisier photometry due to source blending. Fluxes in
the PACS maps were measured by fitting scaled PSFs at each object position, as in \citet{Magnelli_2009}. In the
case of the longer wavelength HerMES data, photometry was performed on all sources simultaneously, with
the 24\,$\micron$ catalogue being used to provide reliable deblending, using a slightly modified version of the
method described in \citet{Roseboom_2010}. The changes to the method are described in \citet{Roseboom_2011};
briefly, a global (rather than local) background estimate was used in producing the catalogues used in this
work, and a different (and faster) model selection algorithm was used in the fitting procedure. Using this
deblending method, reliable fluxes can be extracted close to the formal $\approx 4-5$\,mJy SPIRE confusion noise
\citep[measured after a 3$\sigma_{\rm conf}$ source cut, where $\sigma_{\rm conf}$ is the confusion noise measured 
without this cut,][]{Nguyen_2010}. The 24\,$\micron$ prior positional information reduces the impact of confusion
noise, and so the approximate $3\sigma$ limit for the SPIRE catalogue at 250\,$\micron$ used in this work is
$\approx 9$\,mJy in both fields. We use this catalogue to investigate the properties of GNS galaxies detected
at 250\,$\micron$ in Section~\ref{s_detections}.

In Section~\ref{s_stacking} we present a stacking analysis of GNS galaxies in bins of stellar mass, and we
use data from other infrared surveys to broaden the wavelength coverage outside of the \textit{Herschel}
bands. In both the GOODS-N and GOODS-S fields we use \textit{Spitzer} MIPS 24\,$\micron$ maps, taken from the
Far Infrared Deep Extragalactic Legacy Survey (FIDEL DR2; PI: Mark Dickinson; for GOODS-S) and the
GOODS-\textit{Spitzer} survey (for GOODS-N). In addition, in GOODS-N we make use of the combined AzTEC/MAMBO
1160\,$\micron$ map of \citet{Penner_2011}, while in GOODS-S we use the 870\,$\micron$ LABOCA map from LESS
\citep{Weiss_2009}. To simplify the stacking analysis, the MIPS and PACS maps (in surface brightness units)
are cross-correlated with the appropriate area normalised point spread function such that each pixel in the
resulting map represents the maximum likelihood flux density (in Jy) of an isolated point source at that
position. For the publicly available AzTEC/MAMBO and LESS maps, this operation has
already been performed.

\section{Properties of SPIRE detected GNS galaxies}
\label{s_detections}

\subsection{Cross-matching}
\label{s_crossMatching}

We cross-match the GNS catalogue with the HerMES/PEP catalogue using a simple 2$\arcsec$ matching radius.
Since the HerMES/PEP catalogue was extracted using MIPS 24\,$\micron$ prior positions, a small matching radius,
appropriate to the astrometric accuracy achievable with MIPS at 24\,$\micron$, can be
used \citep[e.g.][]{Bai_2007}. We select robust detections at 250\,$\micron$ from the catalogue
using cuts of $S_{250} > 3 \times \Delta S_{250}$, where $\Delta S_{250}$ is the flux uncertainty (including
confusion noise), i.e. $S_{250} > 8-9$\,mJy (see Section~\ref{s_IRData}), and $\chi^2 < 5$ \citep[i.e. the
goodness of fit of the source solution within the neighbourhood of the source, see][]{Roseboom_2010}. We find
that a total of 22 GNS galaxies with $1.5 < z < 3$ and $\log (M_*/{\rm M_{\sun}}) > 9.5$ are matched across both
the GOODS-N and GOODS-S fields; this corresponds to $\approx 2.5$ per cent of the GNS sample within these stellar
mass and redshift cuts.  We note that if we repeat the selection at 350\,$\micron$, we obtain a sample 
of 14 objects, only 1 of which is not in common with the 250\,$\micron$ selected sample. This additional 
source is ID~283 in the GNS catalog, and has photometric redshift $z_p = 1.55 \pm 0.15$ and stellar mass
$\log (M_*/{\rm M_{\sun}}) \approx 10.6$.

Fig.~\ref{f_postageStamps} shows $10\arcsec \times 10\arcsec$ NICMOS F160W postage stamp images centred on
each detected GNS galaxy, with the position of the HerMES source and the 2$\arcsec$ matching radius
indicated. In almost all cases each GNS galaxy is unambiguously identified with the HerMES source; there
are only two cases (IDs 4180 and 5310) where two galaxies of similar brightness are located within the
matching circle. We estimated the fraction of potentially spurious matches by randomising the positions of 
the sub-mm sources and repeating the cross matching procedure 1000 times. We found a mean number of $3 \pm 2$
of the 250\,$\micron$ sources were randomly associated with GNS galaxies in this test (where the uncertainty is the 
standard deviation). This can be treated as an upper limit, as it assumes no correlation between objects 
detected in the sub-mm and near-IR - and so the real fraction of spurious matches is likely to be lower.

Table~\ref{t_detectedSourceProperties} lists the properties (redshift, stellar mass, rest-frame colour) and
flux densities of the individual detected sources. The median redshift of the detected objects is $z = 2.02$,
and the median stellar mass of the detections is $\log (M_*/{\rm M_{\sun}}) = 10.8$. We note that in 
comparison to the bulk of the GNS sample (Section~\ref{s_sample}), these objects typically have lower
photometric redshift probabilities, with median $P = 61$.

\begin{figure}
\includegraphics[width=9cm]{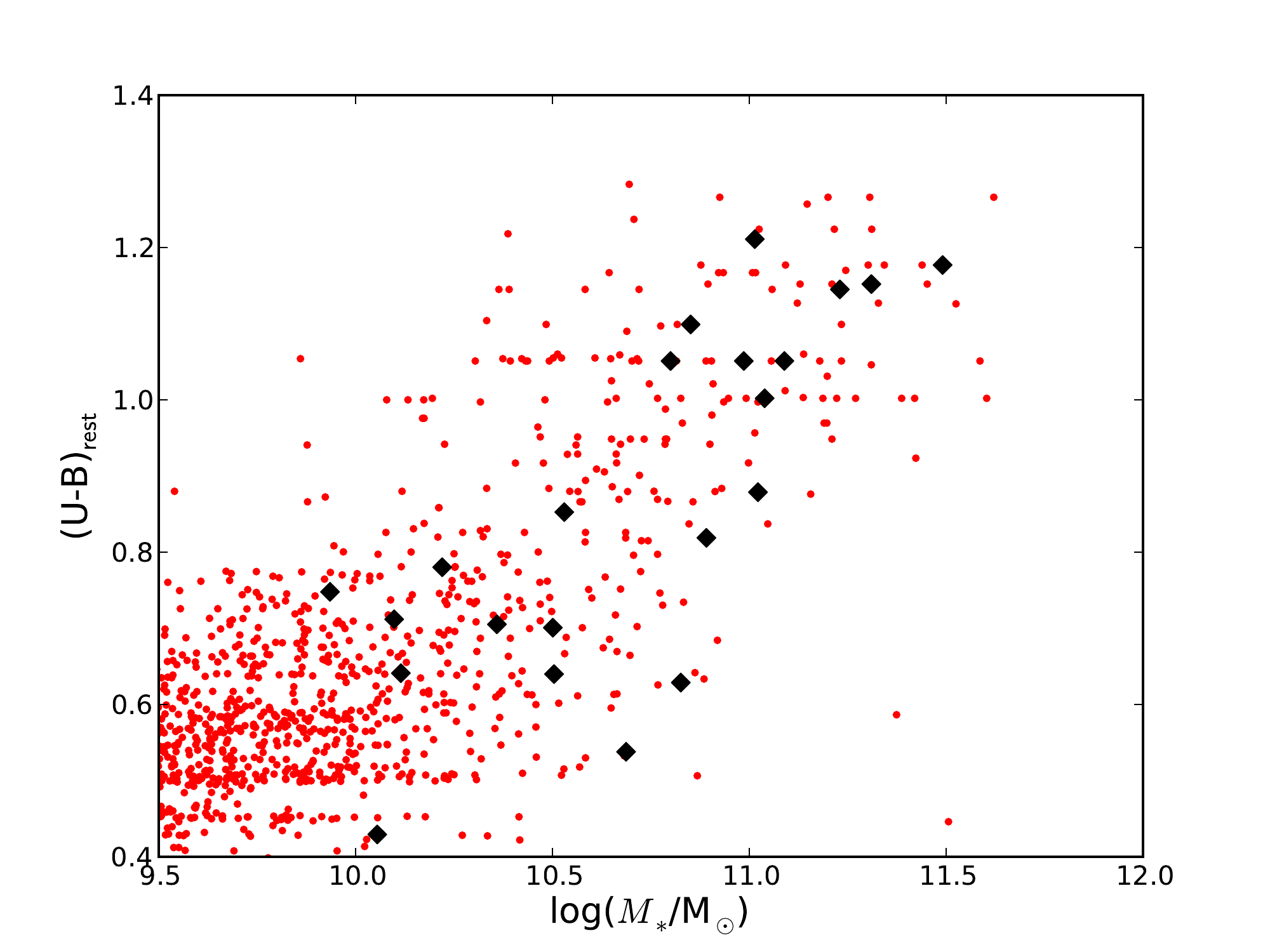}
\caption{Distribution of $1.5 < z < 3$ GNS galaxies with $\log (M_*/{\rm M_{\sun}}) > 9.5$ in the ($U-B$)$_{\rm rest}$
colour--stellar mass plane (small red dots). The large black diamonds indicate the objects detected at
250\,$\micron$ in HerMES. The typical uncertainty in the GNS stellar mass estimates is $\sim0.2$\,dex, while the
typical uncertainty in ($U-B$)$_{\rm rest}$ is 0.15 mag \citep[see][]{Conselice_2011}.}
\label{f_UBStellarMass}
\end{figure}

\begin{figure}
\includegraphics[width=8.6cm]{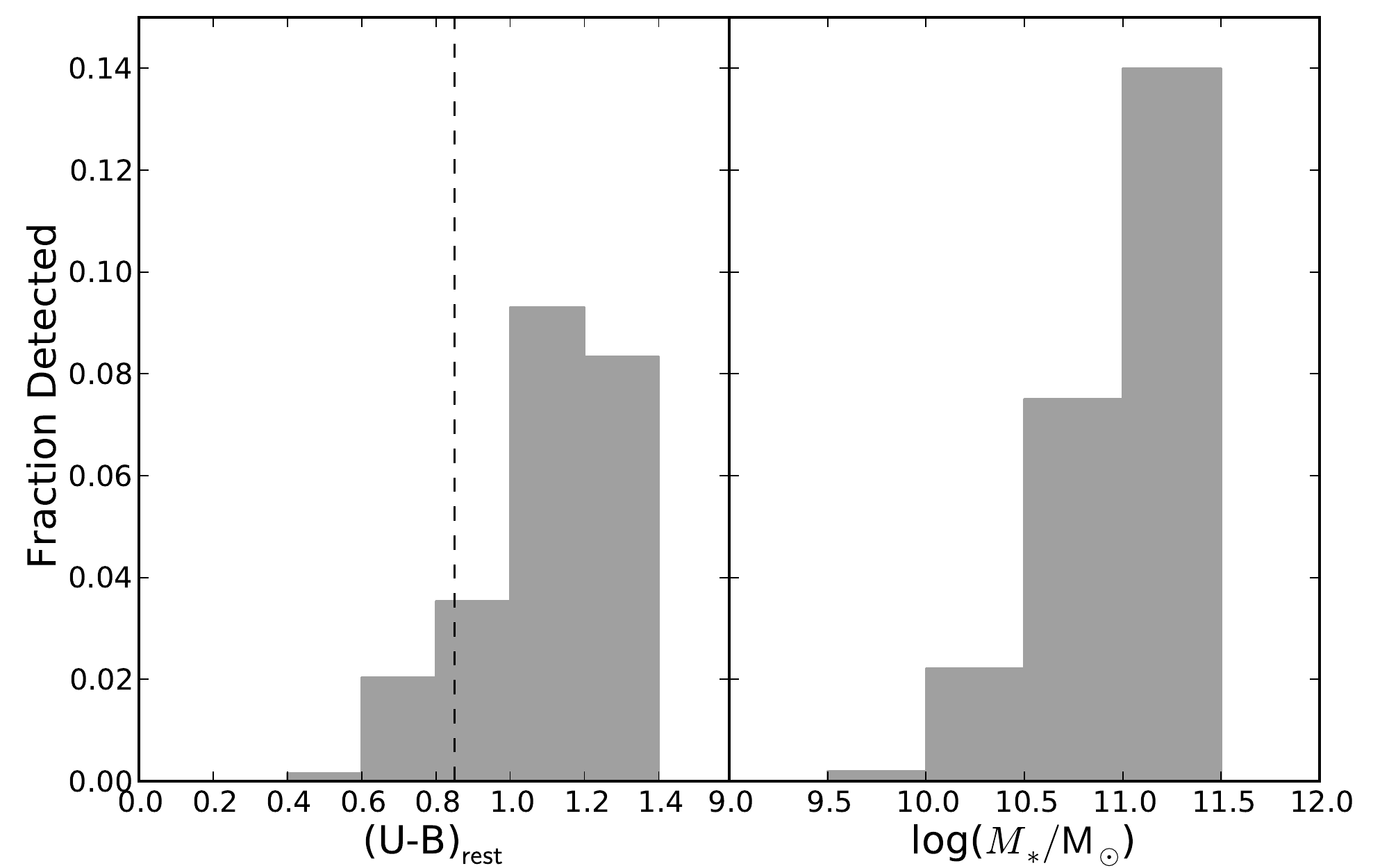}
\caption{Fraction of GNS galaxies with $\log (M_*/{\rm M_{\sun}}) > 9.5$ and $1.5 < z < 3$ detected with SNR $ > 3$ at
250\,$\micron$ as functions of rest-frame ($U-B$)$_{\rm rest}$ colour (left) and stellar mass (right). Clearly,
massive galaxies with redder colours are preferentially detected. For comparison, the rest-frame colour 
separation between quiescent and actively star forming galaxies adopted by \citet{Kriek_2009} is at 
($U-B$)$_{\rm rest}$ $ = 0.85$ (dashed line).}
\label{f_UBFraction}
\end{figure}

\begin{figure}
\includegraphics[width=8.5cm]{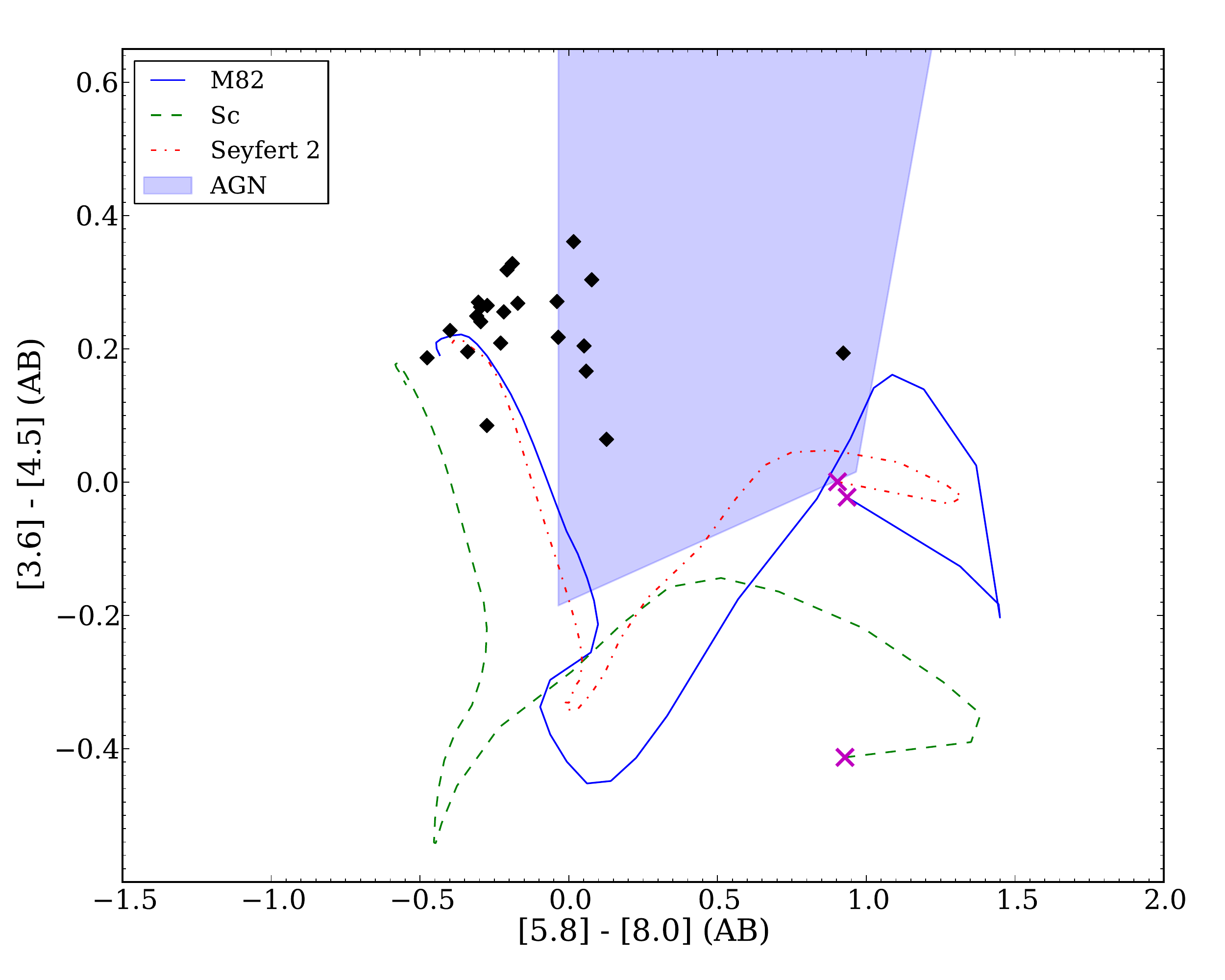}
\caption{IRAC colour--colour plot of GNS galaxies detected in HerMES. Overplotted are non-evolving tracks
of various spectral templates as they are redshifted from $z=0$ to $z=2$ (see legend; the crosses indicate the
$z=0$ end of each track), taken from the library of \citet{Polletta_2007}. The colours of most of the objects
are not consistent with those expected of Type I QSOs (shown by the shaded area marked `AGN' in the legend),
and are more similar to those expected of star forming galaxies at this redshift.}
\label{f_IRACColCol}
\end{figure}

Fig.~\ref{f_UBStellarMass} shows the location of the detected objects in the ($U-B$) colour--stellar mass plane.
Clearly, relatively more massive galaxies with red rest-frame ($U-B$) colours are detected, as shown in
Fig.~\ref{f_UBFraction}. We find that roughly 13 per cent of the sample with $\log(M_*/{\rm M_{\sun}}) > 11$ 
and ($U-B$)$_{\rm rest} > 0.85$ (the fiducial colour criterion adopted for dividing quiescent and star forming
galaxies in \citealt{Kriek_2009}) are detected at 250\,$\micron$. Given their far-infrared flux densities, 
these objects are clearly not quiescent, and we expect them to have high dust masses and high star formation 
rates, with their red colours being as a result of dust extinction. However, it is possible that the dominant
origin of the IR emission is hot dust associated with AGN, rather than star formation, although this is not 
likely: e.g. \citet{Symeonidis_2010} found that all of their 70\,$\micron$
selected galaxy sample were primarily powered by star formation, Although X-ray AGN were
removed from the sample at the outset (Section~\ref{s_sample}), we checked for additional AGN using colours in
the \textit{Spitzer} IRAC bands \citep{Stern_2005}, using data from the GOODS \textit{Spitzer} Legacy program 
\citep{Dickinson_2003}. Fig.~\ref{f_IRACColCol} shows the
[3.6]--[4.5], [5.8]--[8.0] colour--colour plot of the 250\,$\micron$ detected GNS galaxies. We find that six
objects fall within the region typically occupied by AGN. We do not remove these objects from the sample, as
some studies have shown that AGN mainly contribute to the IR flux at wavelengths $< 20$\,$\micron$
\citep[][see also \citealt{Hatziminaoglou_2010}]{Netzer_2007, Mullaney_2011}; we will instead note these 
objects in the following analysis (see also Section~\ref{s_CIGALE} below).

We note that it is possible that the presence of either an AGN or starburst may lead to the 
stellar masses of some of the detected sources being overestimated. Other studies, which explicitly correct for
the effect of power law emission from AGN, find that neglecting such corrections can lead to differences of
10-25 per cent in stellar mass estimates of SMGs \citep[e.g.][]{Hainline_2011}. We show in Section~\ref{s_CIGALE}
below that more sophisticated SED modelling, using rather different assumptions to those used in deriving the 
GNS stellar masses, verifies that the 250\,$\micron$ detected GNS galaxies are genuinely massive systems. 
\citep[see also the discussion concerning stellar mass estimates of AGN hosting GNS galaxies in][]{Bluck_2011}.

\begin{table*}
\caption{Properties of $1.5 < z < 3.0$ GNS galaxies detected at 250~$\micron$ with $S/N > 3$. Flux densities ($S_\lambda$) are in mJy, and only wavelengths in common between both GOODS-N and GOOD-S are shown.
The error bars on photometric redshifts (we do not show error bars on objects with spectroscopic redshifts, marked
with $\dagger$) and stellar mass estimates are statistical only, and the typical uncertainty in ($U-B$)$_{\rm rest}$ is 0.15 mag \citep[see][for details]{Conselice_2011}.}
\label{t_detectedSourceProperties}
\begin{tabular}{|c|c|c|c|c|c|c|c|c|c|}
\hline
    GNS ID & $z$ & log $M_*$ & ($U-B$)$_{\rm rest}$ & $S_{24}$ & $S_{100}$ & $S_{160}$ & $S_{250}$ & $S_{350}$ & $S_{500}$\\
\hline
77\phantom{0}   &$2.33 \pm 0.20$ & $11.09 \pm 0.01$           & 1.05 & $0.332 \pm 0.007$& \phantom{0}\nodata        & \phantom{0}\nodata        & $20.7 \pm 3.1$            & $17.3 \pm 4.1$            & $15.1 \pm 4.4$\\
895$\star$      &2.08$\dagger$   & $10.05 \pm 0.01$           & 0.43 & $0.080 \pm 0.005$& \phantom{0}\nodata        & \phantom{0}\nodata        & $12.1 \pm 3.1$            & \phantom{0}$5.8 \pm 4.3$  & $11.9 \pm 4.3$\\
1150\phantom{0} &$1.86 \pm 0.17$ & $10.10 \pm 0.17$           & 0.71 & $0.256 \pm 0.006$& \phantom{0}\nodata        & \phantom{0}$5.9 \pm 1.7$  & $10.5 \pm 3.1$            & $12.4 \pm 4.1$            & \phantom{0}\nodata\\
1394\phantom{0} &$2.29 \pm 0.20$ & $11.49 \pm 0.11$           & 1.18 & $0.178 \pm 0.006$& \phantom{0}\nodata        & \phantom{0}$7.7 \pm 2.4$  & $19.3 \pm 3.1$            & $17.5 \pm 4.0$            & \phantom{0}$3.3 \pm 3.9$\\
1754\phantom{0} &$1.72 \pm 0.16$ & $10.22 \pm 0.18$           & 0.78 & $0.169 \pm 0.006$& \phantom{0}\nodata        & \phantom{0}\nodata        & \phantom{0}$9.7 \pm 3.1$  & $10.7 \pm 4.0$            & \phantom{0}$3.8 \pm 4.0$\\
2138$\star$     &$2.03 \pm 0.18$ & \phantom{0}$9.94 \pm 0.18$ & 0.75 & $0.117 \pm 0.007$& \phantom{0}\nodata        & \phantom{0}\nodata        & $10.9 \pm 3.1$            & \phantom{0}$7.9 \pm 4.0$  & \phantom{0}\nodata\\
2411\phantom{0} &$2.08 \pm 0.19$ & $11.04 \pm 0.07$           & 1.00 & $0.298 \pm 0.006$& \phantom{0}\nodata        & \phantom{0}$6.5 \pm 1.8$  & \phantom{0}$9.3 \pm 3.1$  & \phantom{0}$9.4 \pm 4.1$  & \phantom{0}$0.9 \pm 4.0$\\
3511\phantom{0} &$2.35 \pm 0.20$ & $10.80 \pm 0.07$           & 1.05 & $0.086 \pm 0.008$& \phantom{0}\nodata        & \phantom{0}\nodata        & $10.0 \pm 3.1$            & \phantom{0}$5.0 \pm 4.2$  & \phantom{0}\nodata\\
3966$\star$     &2.46$\dagger$   & $10.69 \pm 0.10$           & 0.54 & $0.142 \pm 0.007$& \phantom{0}\nodata        & \phantom{0}\nodata        & $11.5 \pm 3.1$            & $12.8 \pm 4.0$            & $13.1 \pm 4.2$\\
4180$\star$     &2.00$\dagger$   & $10.36 \pm 0.13$           & 0.71 & $1.218 \pm 0.012$& $11.5 \pm 1.0$            & \phantom{0}\nodata        & $23.1 \pm 3.1$            & $24.8 \pm 4.1$            & $10.8 \pm 4.0$\\
4754\phantom{0} &$2.40 \pm 0.20$ & $11.31 \pm 0.16$           & 1.15 & $0.440 \pm 0.006$& \phantom{0}$2.6 \pm 0.3$  & \phantom{0}$6.6 \pm 0.6$  & $12.1 \pm 2.6$            & \phantom{0}$4.6 \pm 3.4$  & \phantom{0}$2.4 \pm 4.2$\\
5040\phantom{0} &$1.71 \pm 0.16$ & $10.12 \pm 0.15$           & 0.64 & $0.220 \pm 0.006$& \phantom{0}$2.1 \pm 0.3$  & \phantom{0}$7.6 \pm 0.8$  & $13.9 \pm 2.6$            & $11.6 \pm 3.3$            & \phantom{0}$8.0 \pm 4.2$\\
5282\phantom{0} &$1.64 \pm 0.16$ & $11.02 \pm 0.06$           & 0.88 & $0.456 \pm 0.005$& \phantom{0}$2.6 \pm 0.3$  & \phantom{0}$7.8 \pm 0.4$  & $13.2 \pm 2.6$            & $14.9 \pm 3.2$            & \phantom{0}$3.2 \pm 4.3$\\
5306\phantom{0} &$1.55 \pm 0.15$ & $10.85 \pm 0.03$           & 1.10 & $0.328 \pm 0.005$& \phantom{0}$2.9 \pm 0.4$  & \phantom{0}$7.0 \pm 0.5$  & $18.2 \pm 2.6$            & $15.0 \pm 3.7$            & $24.8 \pm 4.4$\\
5310\phantom{0} &1.82$\dagger$   & $10.89 \pm 0.17$           & 0.82 & $0.237 \pm 0.005$& \phantom{0}$4.7 \pm 0.4$  & $14.4 \pm 0.5$            & $16.0 \pm 2.6$            & $14.4 \pm 3.5$            & \phantom{0}\nodata\\
5853$\star$     &2.41$\dagger$   & $10.50 \pm 0.13$           & 0.64 & $0.166 \pm 0.003$& \nodata                   & \phantom{0}\nodata        & \phantom{0}$8.2 \pm 2.6$  & $16.6 \pm 3.2$            & \phantom{0}$8.2 \pm 4.4$\\
5918\phantom{0} &1.98$\dagger$   & $10.83 \pm 0.11$           & 0.63 & $0.277 \pm 0.004$& \phantom{0}$3.5 \pm 0.6$  & \phantom{0}$7.5 \pm 0.9$  & $11.6 \pm 2.6$            & $14.6 \pm 3.5$            & $20.8 \pm 4.3$\\
6081\phantom{0} &$1.64 \pm 0.16$ & $10.53 \pm 0.14$           & 0.85 & $0.199 \pm 0.004$& \phantom{0}$2.0 \pm 0.3$  & \phantom{0}$4.8 \pm 0.7$  & \phantom{0}$8.4 \pm 2.6$  & \phantom{0}$5.9 \pm 3.1$  & \phantom{0}\nodata\\
6160$\star$     &$2.31 \pm 0.20$ & $10.50 \pm 0.13$           & 0.70 & $0.234 \pm 0.005$& \phantom{0}$2.1 \pm 0.4$  & \phantom{0}$4.3 \pm 0.6$  & \phantom{0}$8.7 \pm 2.6$  & \phantom{0}$8.4 \pm 3.2$  & $16.6 \pm 4.2$\\
6220\phantom{0} &$1.72 \pm 0.16$ & $11.01 \pm 0.19$           & 1.21 & $0.167 \pm 0.003$& \nodata                   & \phantom{0}$7.0 \pm 1.5$  & $18.4 \pm 2.6$            & $15.2 \pm 3.2$            & \phantom{0}$7.0 \pm 4.3$\\
7475\phantom{0} &1.61$\dagger$   & $11.23 \pm 0.06$           & 1.15 & $0.191 \pm 0.004$& \phantom{0}$1.8 \pm 0.4$  & \phantom{0}$4.6 \pm 0.8$  & $12.9 \pm 2.6$            & \phantom{0}$7.1 \pm 3.3$  & \phantom{0}\nodata\\
7970\phantom{0} &$2.54 \pm 0.21$ & $10.99 \pm 0.13$           & 1.05 & $0.264 \pm 0.004$& \phantom{0}$1.7 \pm 0.5$  & \phantom{0}$4.8 \pm 1.1$  & $11.8 \pm 2.6$            & $10.3 \pm 3.2$            & \phantom{0}$4.6 \pm 4.3$\\
\hline
\end{tabular}
\medskip{}
\\
\raggedright{$\star$ = IRAC colours of this object indicates AGN may be present (see Fig. 5)}\\
\raggedright{$\dagger$ = spectroscopic redshift (taken from the compilations by \citealt{Barger_2008} and \citealt{Wuyts_2008})\\}
\end{table*}   
    
\subsection{SED fitting}
\label{s_fitting}
To estimate $L_{\rm IR}$ and SFR for the SPIRE detected GNS galaxies, we fit their far-IR SEDs using
a modified blackbody \citep[e.g.][]{Hildebrand_1983, Blain_2003} of the form,
\begin{equation}
\label{eq_blackbody}
S_\nu = A \nu^\beta B(\nu, T_{\rm dust}),
\end{equation}
where $B(\nu, T_{\rm dust})$ is the Planck function, $A$ is the amplitude, and $\beta$ is the emissivity index 
(fixed to $\beta = 1.5$). In addition, the Wien tail is replaced with a power
law of the form $S_\nu \propto \nu^{-\alpha}$, with $\alpha = -2$ \citep{Blain_2003}. We also fit the SEDs
using the templates of \citet[][CE01 hereafter]{CharyElbaz_2001}, as a consistency check on our results. 
\defcitealias{CharyElbaz_2001}{CE01}

We fit the SEDs using $\chi^2$ minimisation, allowing the dust temperature
to vary in the range $10 - 70$\,K. We ignore the 24\,$\micron$ flux densities when fitting the SEDs using models of the
form of equation~\ref{eq_blackbody}, since at $z > 1.5$ we do not expect the modified blackbody model to be a
reasonable description of the SED at this wavelength in the observed frame. However, we do include the
24\,$\micron$ fluxes when fitting to the \citetalias{CharyElbaz_2001} templates, as these include the
contribution from polyaromatic hydrocarbon (PAH) features. Note that we include SED points with SNR$< 3$
in the fitting - given the requirement of a 24\,$\micron$ detection and prior position, so long as the 
uncertainties on these points are accurately estimated, then the additional information they provide should 
help to better constrain the SED than either neglecting these points, or replacing them with 3$\sigma$ upper 
limits. We comment on the effect of this on our results in Section~\ref{s_detectionsSFR}.

We derive the total
($8-1000$\,$\micron$) IR luminosity ($L_{\rm IR}$) from the amplitude of the best fitting model, and convert
this to a SFR, assuming that the \citet{Kennicutt_1998} law holds at this redshift,
\begin{equation}
\label{eq_kennicutt}
{\rm SFR_{IR}} \ ({\rm M_{\sun}}~{\rm yr^{-1}}) = (4.5 \times 10^{-44}) \times {L_{\rm IR}} \ ({\rm
erg~s^{-1}}),
\end{equation}
defined with respect to a \citet{Salpeter_1955} IMF. We therefore apply a correction of $-0.23$ dex to
SFRs estimated using equation~\ref{eq_kennicutt} to account for the \citet{Chabrier_2003} IMF assumed in this
work \citep[see e.g.][]{Kriek_2009}.

We also estimate dust masses during the SED fitting, using the method of \citet[][see also
\citealt{Dunne_2000} and references therein]{Dunne_2011}, i.e.,
\begin{equation}
\label{eq_dustMass}
M_{\rm dust} = \frac{S_{\rm 250} D_{\rm L}^2 K}{\kappa_{250}B(\nu, T_{\rm dust})},
\end{equation}
where $S_{\rm 250}$ is the flux density at 250\,$\micron$ in the observed frame, $K$ is the $K$-correction to 
rest-frame 250\,$\micron$,
$D_{\rm L}$ is the luminosity distance, and $\kappa_{250}$ is the dust mass absorption coefficient, taken to be 
0.89\,m$^2$\,kg$^{-1}$ as in \citet{Dunne_2011}. There are many caveats for the dust mass estimates obtained in
this way, such as: the uncertainty in the value of $\kappa_{250}$; the fact that equation~\ref{eq_dustMass}
can underestimate the true dust mass due to the presence of warm dust in galaxies being neglected in the
modified blackbody model (equation~\ref{eq_blackbody}); and the large $K$-correction to the redshift range of
our study. Although the absolute values of $M_{\rm dust}$ are highly uncertain, we use the relative values
obtained by this method to give an indication of the relation of $M_{\rm dust}$ with $M_*$, assuming that the
dust properties are similar in galaxies of different stellar mass in our redshift range of interest (see
Section~\ref{s_stackingResults}).

We estimate errors on the parameters derived from the SED fits using Monte-Carlo simulations. For each
observed SED we generate 1000 random realisations, assuming that the errors on the fluxes are
Gaussian. For objects with only photometric redshifts, we simultaneously randomise the redshift of the fitted
model SED according to the scatter of $\sigma_z = 0.06$ measured by \citet{Gruetzbauch_2011}. We adopt the 68.3 percentile range
from the distribution of parameter values obtained from the random realisations as the corresponding $\pm 1\sigma$
uncertainty.

\subsection{Results}
\label{s_detectionsSFR}

\subsubsection{Star formation}

The SED fitting shows that the GNS galaxies individually detected in HerMES are ULIRGs, spanning the
range $11.9 < $~log~$L_{\rm IR} (\rm L_{\sun}) < 12.9$, with mean log~$L_{\rm IR} = 12.36 \pm 0.05$~$L_\odot$,
where the quoted uncertainty is the standard error on the mean. We estimate total SFRs for
these galaxies under the assumption that this corresponds to the sum of the SFR derived from the far-IR SEDs 
and the UV-based unobscured SFR measurements from \citet{Bauer_2011}. We find that the mean total SFR for 
these galaxies is SFR$_{\rm IR+UV} = 280 \pm 40$~$\rm M_{\sun}$~yr$^{-1}$. Removing the six
galaxies with IRAC colours consistent with AGN has no significant effect: with these objects
excluded, we find SFR$_{\rm IR+UV} = 260 \pm 50$~$\rm M_{\sun}$~yr$^{-1}$. This is a factor of $> 2$ larger
than the mean UV-slope extinction corrected SFR estimates from \citet{Bauer_2011} for these same galaxies, i.e. SFR$_{\rm
UV,corr} = 120 \pm 30$~$\rm M_{\sun}$~yr$^{-1}$. We obtain results within $<1\sigma$ of these values for all 
of these properties if we take into account the fraction of potential spurious matches 
(Section~\ref{s_crossMatching}) in a Monte-Carlo fashion. 

We checked the sensitivity of these estimates to the adopted sub-mm selection criteria. We find 
consistent results for the smaller sample of 8 galaxies detected with SNR $ > 5$ at 250\,$\micron$ 
(mean log~$L_{\rm IR} = 12.39 \pm 0.09$~$L_\odot$, mean 
SFR$_{\rm IR+UV} = 290 \pm 60$~$\rm M_{\sun}$~yr$^{-1}$), and for the sample of 14 galaxies detected at 
SNR $> 3$ at 350\,$\micron$ (mean log~$L_{\rm IR} = 12.34 \pm 0.07$~$L_\odot$, mean
SFR$_{\rm IR+UV} = 260 \pm 40$~$\rm M_{\sun}$~yr$^{-1}$). We also checked the effect of including SED points
with SNR $< 3 $ in the fits (see Section~\ref{s_fitting}) - replacing them with $3\sigma$ upper
limits, we obtain mean log~$L_{\rm IR} = 12.40 \pm 0.05$~$L_\odot$, with corresponding mean 
SFR$_{\rm IR+UV} = 300 \pm 40$~$\rm M_{\sun}$~yr$^{-1}$, for the whole sample of 22 galaxies.

Dividing the sample by rest frame colour, we see no evidence for different 
IR properties for galaxies detected at 250\,$\micron$ with red or blue colours, although of course the sample is very small. We find 
mean $\log L_{\rm IR} = 12.33 \pm 0.09$ (SFR$_{\rm IR+UV} = 270 \pm 60$~$\rm M_{\sun}$~yr$^{-1}$) for the 11
galaxies with ($U-B$)$_{\rm rest} > 0.85$, and mean $\log L_{\rm IR} = 12.34 \pm 0.06$ 
(SFR$_{\rm IR+UV} = 260 \pm 40$$\rm M_{\sun}$~yr$^{-1}$) for the 11 galaxies with ($U-B$)$_{\rm rest} < 0.85$.

We conclude that SFR$_{\rm IR+UV}$ is significantly higher than SFR$_{\rm
UV,corr}$ for our sample. \citet{Wuyts_2011} also found that SFR$_{\rm UV,corr}$ is underestimated compared to SFR$_{\rm IR+UV}$ for galaxies with similar 
total star formation rates and redshifts to our sample.
However, several other recent studies find the reverse situation. For example,
\citet{Murphy_2011} observed a sample of $0.66 < z < 2.6$ 24\,$\micron$ selected sources with additional
70\,$\micron$ photometry, and found that their measurements of SFR$_{\rm UV,corr}$ are a factor of $> 2$
\textit{higher} than SFR$_{\rm IR+UV}$. They concluded that the dust corrections applied to their sample (from
the \citealt{Meurer_1999} relation) were overestimated for many objects. \citet{Nordon_2010} found similar
results from a study using PACS observations of massive galaxies at $1.5 < z < 2.5$ in GOODS-N, finding
SFR$_{\rm UV,corr}$ is overestimated by a factor of about 2 for galaxies with SFR$_{\rm UV} >
40$~$\rm M_{\sun}$~yr$^{-1}$, assuming a Calzetti UV attenuation law (note however that \citealt{Wuyts_2011} showed that this result 
may in part be driven by the relatively bright $K_s < 22$ limit adopted in \citealt{Nordon_2010}). \citet{Buat_2010} reached similar
conclusions from a study of 250\,$\micron$ selected $z < 1$ galaxies from HerMES with UV photometry from the
\textit{Galaxy Evolution Explorer} (\textit{GALEX}) satellite. At lower redshift ($z < 0.35$), 
\citet{Wijesinghe_2011} found only a weak correlation with large scatter
between the UV slope ($\beta$) and $L_{\rm IR}/L_{\rm UV}$, which would also lead to overestimated SFR$_{\rm
UV, corr}$. However, for UV selected samples (e.g. Lyman break galaxies; LBGs) which are not ULIRGs, dust
corrections from the local \citet{Meurer_1999} relation appear to be valid at $z \sim 2$
\citep[e.g.][]{Overzier_2011, Reddy_2011}. Reasonable agreement between SFR$_{\rm UV,corr}$ and SFR derived
from stacked radio and 24\,$\micron$ observations is also seen up to $z \sim 3$ for LBGs
\citep{MagdisSFR_2010}.

\begin{figure}
\includegraphics[width=8.5cm]{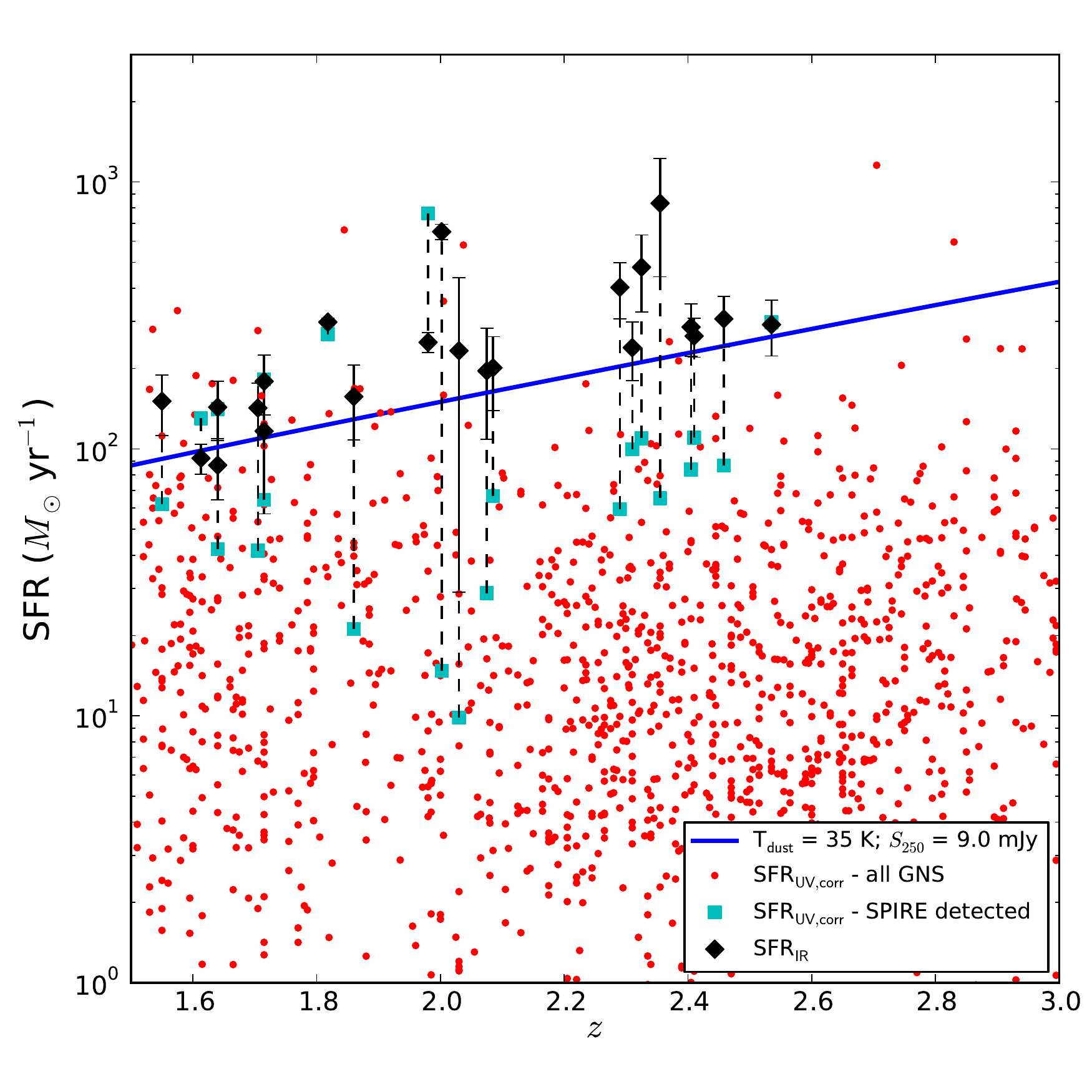}
\caption{Comparison of SFR$_{\rm IR}$ estimated for GNS galaxies detected at 250\,$\micron$ (black diamonds)
with the approximate 3$\sigma$ flux limit as a function of redshift (blue line, estimated assuming a
modified blackbody SED with $T_{\rm dust} = 35$\,K), and the extinction corrected UV estimates (SFR$_{\rm UV,corr}$)
for these same galaxies (cyan squares). The SFR$_{\rm UV,corr}$ values of the entire GNS sample are plotted
for comparison (small red dots). The dashed lines indicate corresponding SFR estimates for a given galaxy.}
\label{f_SFRBias}
\end{figure}

\begin{figure}
\includegraphics[width=8.5cm]{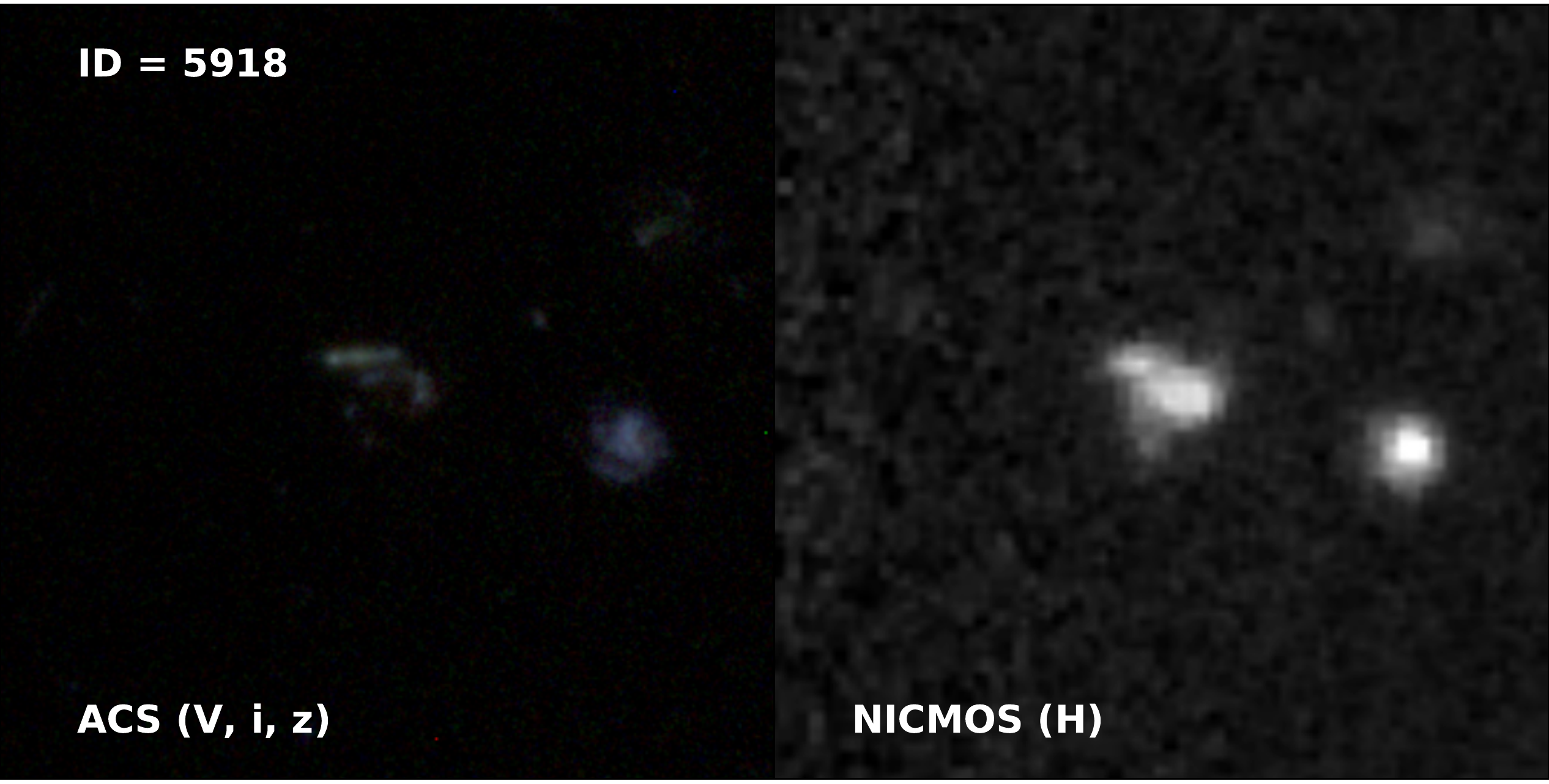}
\caption{ACS ($V$, $i$, $z$) image (10$\arcsec \times$ 10$\arcsec$) of the mulitple component system
ID 5918 (left), the only galaxy in the sample with significantly larger SFR$_{\rm UV, corr}$ than 
SFR$_{\rm IR}$ of the GNS galaxies detected at 250\,$\micron$ (see Fig.~\ref{f_SFRBias}).}
\label{f_outlier}
\end{figure}

\begin{figure}
\includegraphics[width=9cm]{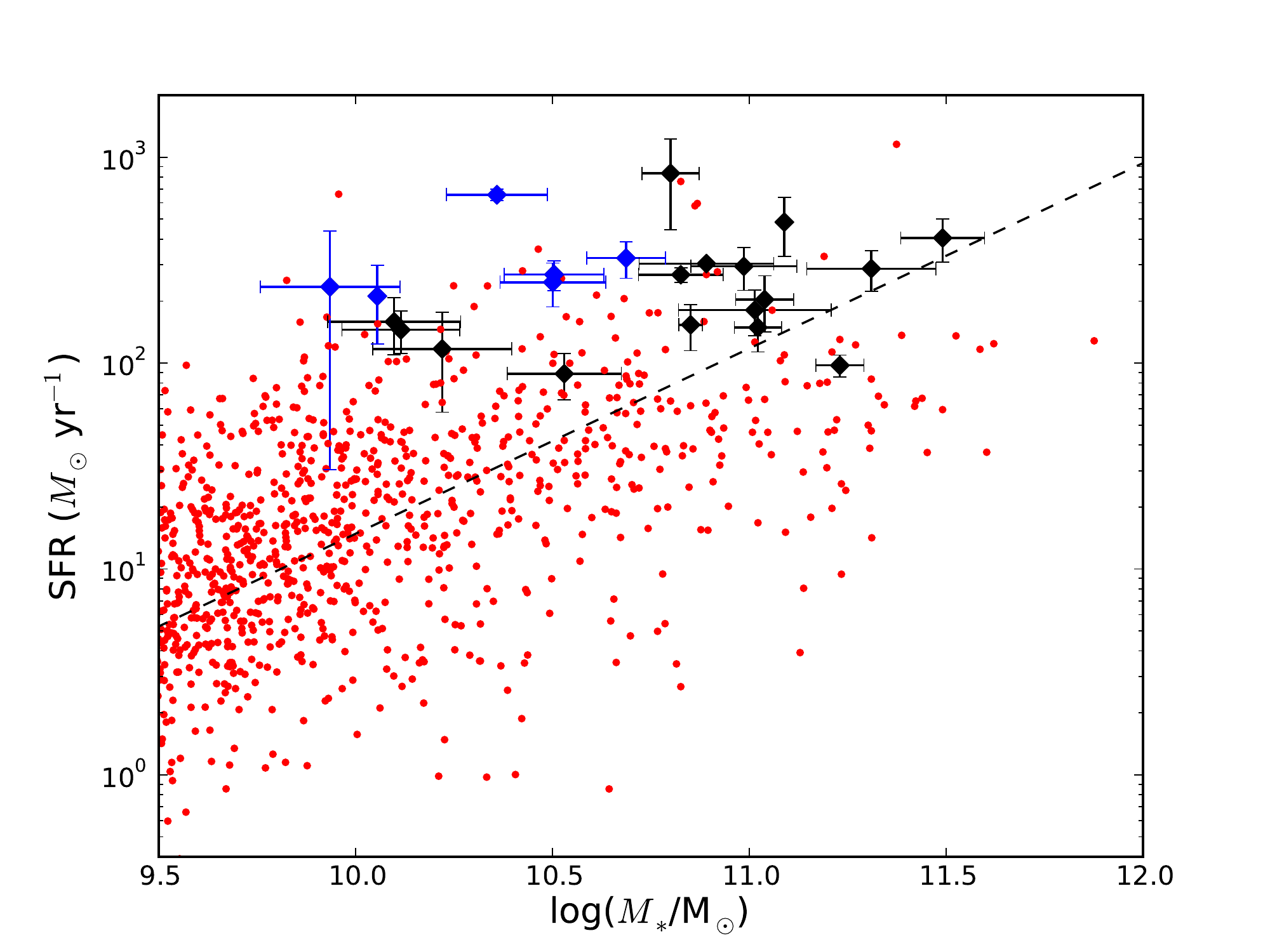}
\caption{Relation between total SFR and $M_*$ for GNS galaxies. The large diamonds represent SPIRE
detected galaxies; those highlighted in blue have IRAC colours consistent with AGN (see
Fig.~\ref{f_IRACColCol}). For these galaxies, the total SFR estimate that we use is SFR$_{\rm IR+UV}$. The small red
points represent the wider GNS sample; in this case, the total SFR estimate is SFR$_{\rm UV,corr}$. The
dashed line is the SFR$_{\rm UV, corr}$--$M_*$ relation measured at $z \sim 2$ by \citet{Daddi_2007I}. Note that
the error bars indicate statistical errors in SFR and $M_*$ only.}
\label{f_SFRMStar}
\end{figure}

We expect large IR-derived SFRs for the galaxies we detect at 250\,$\micron$ given their redshift and the
3$\sigma$ flux limit, which is $\approx 9$\,mJy at 250\,$\micron$ in the GOODS-N field. This leads to a large
Malmquist bias (with some flux-boosting due to the low SNR) in comparison to the UV-derived SFRs, which reach to $\sim 1$~$\rm M_{\sun}$~yr$^{-1}$
\citep{Bauer_2011}. Fig.~\ref{f_SFRBias} shows the SFR$_{\rm IR}$ limit
as a function of redshift for a modified blackbody model SED (equation~\ref{eq_blackbody}) with 
$T_{\rm dust} = 35$\,K, normalised to a 250\,$\micron$ flux density of 9\,mJy. Highlighted in this plot are 
the SFR$_{\rm IR}$ and SFR$_{\rm UV,corr}$ values for the SPIRE-detected galaxies; and clearly in most cases 
SFR$_{\rm UV,corr}$ is much lower than the fiducial SFR$_{\rm IR}$ corresponding to the 250\,$\micron$ flux limit.
This makes the comparison between these two SFR measures for our sample difficult to interpret. There is one
clear exception, where SFR$_{\rm UV, corr}$ is roughly a factor of 3 larger than SFR$_{\rm IR}$ - this is 
ID 5918, which, from inspection of the ACS imaging, seems to be a multiple component merger system, 
with regions of significant unobscured star formation (see Fig.~\ref{f_outlier}). It may be that only one
component of this system is the source of the FIR emission, but it is not possible to determine which using
the current data.

Fig.~\ref{f_SFRMStar} shows the comparison of SFR$_{\rm IR+UV}$ and $M_*$ for the SPIRE-detected GNS
galaxies with the wider GNS sample, where for the latter SFR$_{\rm UV,corr}$ is used as the estimate of the
total SFR. We see that almost all of the SPIRE detected galaxies scatter above the SFR--$M_*$ relation
measured by \citet{Daddi_2007I}, which is as expected given the approximate SFR$_{\rm IR}$ limit shown in
Fig.~\ref{f_SFRBias}. 

\subsubsection{Dust properties}
\label{s_detectionsDust}

\begin{figure}
\includegraphics[width=8.5cm]{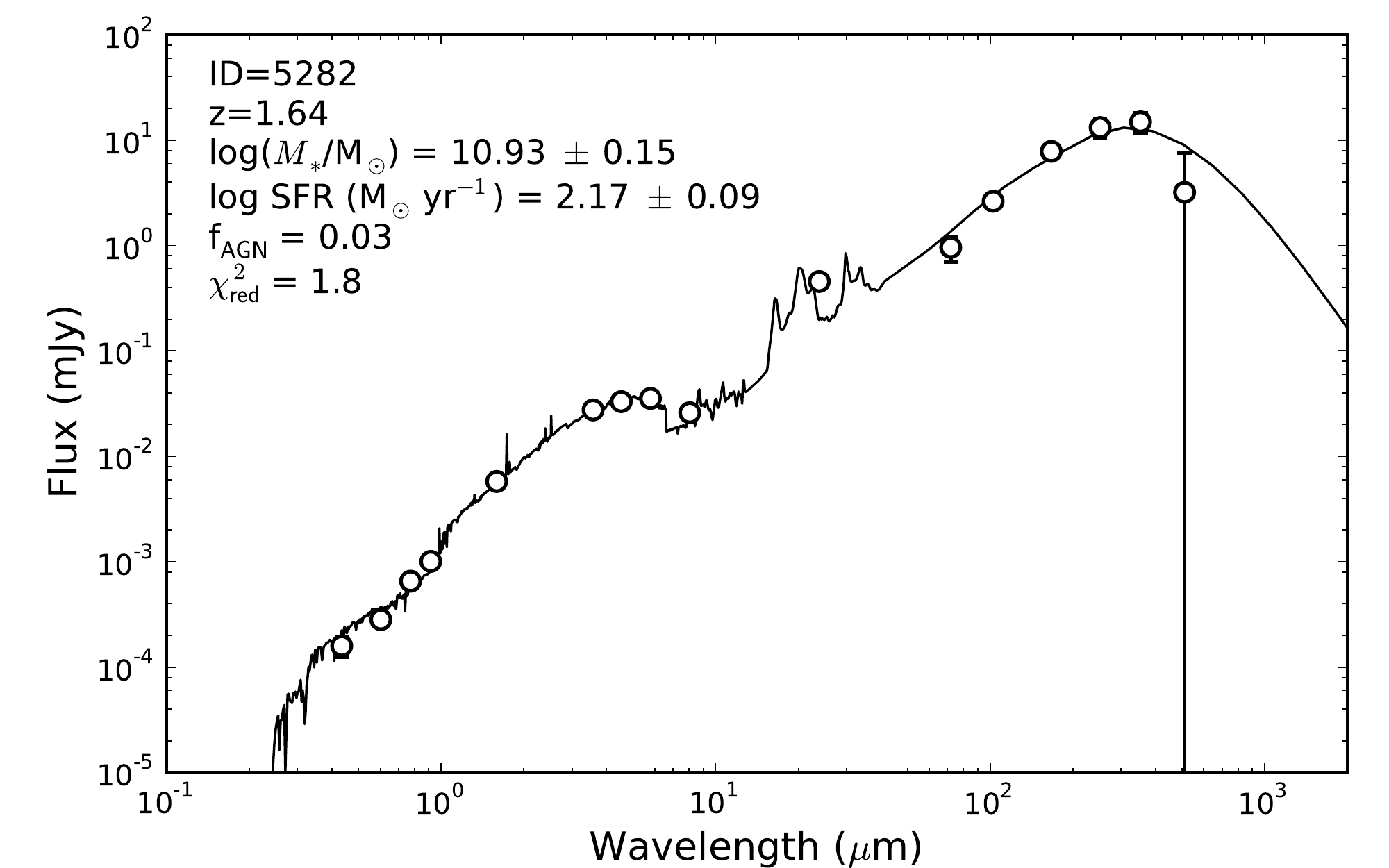}
\includegraphics[width=8.5cm]{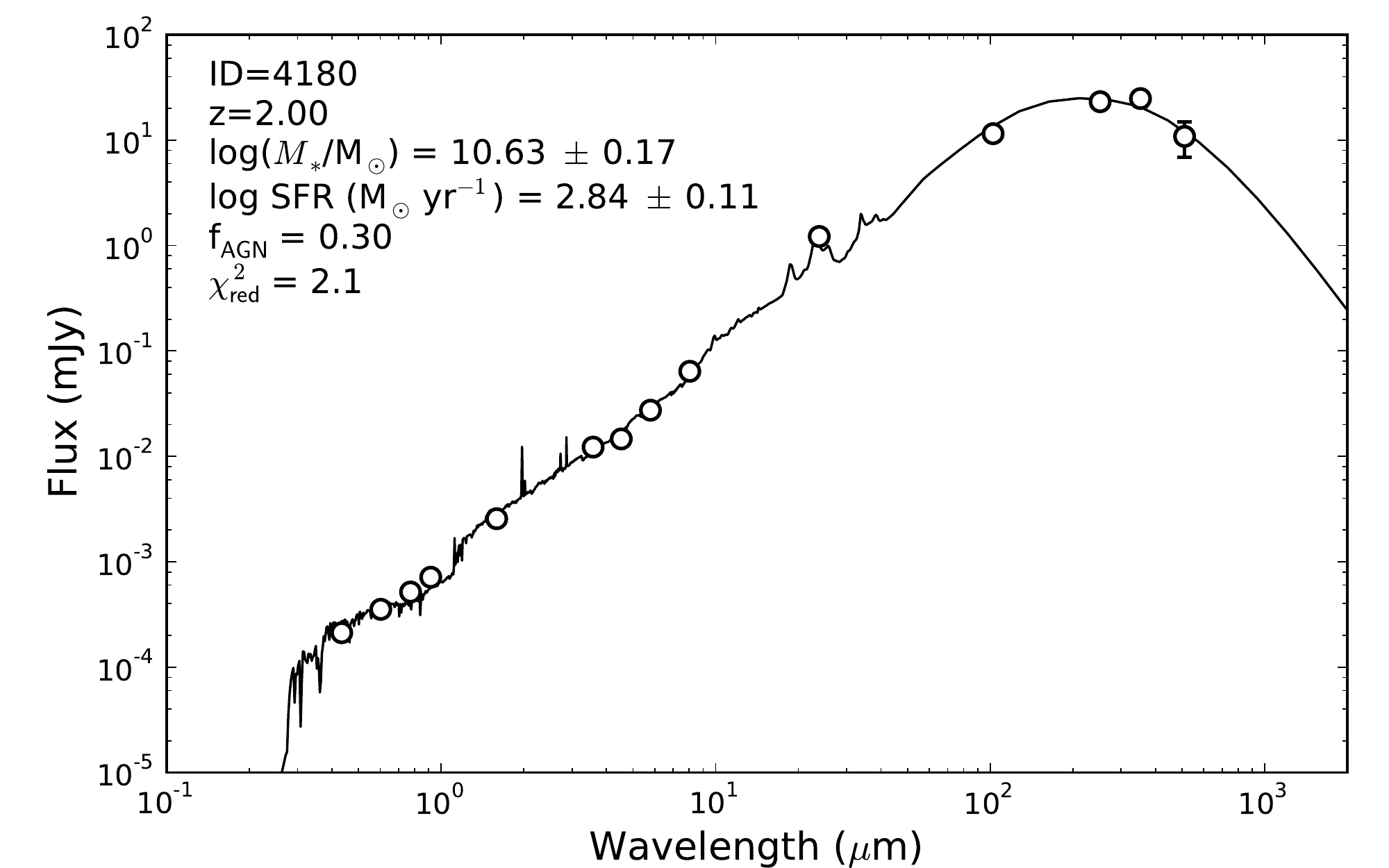}
\caption{Examples of optical-IR SEDs fitted with \texttt{CIGALE}. Note that different underlying assumptions were used with
\texttt{CIGALE} compared to the rest of this work; i.e. the \citet{Maraston_2005} stellar population
models, \citet{Kroupa_2001} IMF, and \citet{DaleHelou_2002} infrared templates were used. The \texttt{CIGALE}
fit results suggest that the bulk of the IR emission is associated with star formation rather than AGN.
Note that the median $\chi^2_{\rm red}$ of the sample is 1.7, so the example fits we show here are representative, 
although we choose to show ID 4180 in particular because it is the object with the largest inferred AGN contribution to the IR luminosity. 
}
\label{f_CIGALE}
\end{figure}

For the 16 galaxies with flux measurements in all SPIRE bands, we find dust temperatures in the range 23--48\,K,
with mean $35\pm 6$\,K (where the quoted uncertainty is the standard deviation). Note however that only 4 of these galaxies
have SNR$> 3$ in all SPIRE bands, and so the individual temperature estimates are poorly constrained, with typical 
statistical uncertainty $\approx 5$\,K. We find that replacing the SNR $< 3$ SED points in the fits with 3$\sigma$
upper limits (see Section~\ref{s_fitting}) gives $T_{\rm dust}$ values for individual galaxies in this subsample
that agree within $<1 \sigma$ of the values obtained when the low SNR SED points are included. For a sample
selected with SNR $> 3$ at 350\,$\micron$, we find mean $T_{\rm dust} = 33 \pm 7$\,K, while for a sample with SNR $> 5$
at 250\,$\micron$, we find mean $T_{\rm dust} = 34 \pm 7$\,K. The single GNS galaxy which is detected at 
SNR $> 3$ at 350\,$\micron$ but is not in our 250\,$\micron$ selected sample (ID 283; see 
Section~\ref{s_crossMatching}) has a slightly lower dust temperature ($T_{\rm dust} = 20 \pm 5$\,K).

The mean temperature we find is somewhat lower than the typical temperature of ULIRGs at $z < 1$ \citep[$T_{\rm dust}
\approx 42$\,K, e.g.][]{Clements_2010, Yang_2007}; although note that $\beta$ is fitted for in the former work, whereas in the latter it is
fixed at $\beta = 1.5$, as we assume here. This is not unexpected given the high
redshift of the sample and the selection at SPIRE wavelengths \citep{Symeonidis_2011}. The dust temperatures
we find are similar to those found for other samples at $z > 1$ \citep{Chapin_2009, Amblard_2010, Chapman_2010, Hwang_2010}. Adopting
$\beta = 2.0$ in the modified blackbody model (equation~\ref{eq_blackbody}) gives mean $T_{\rm dust}$ about
4\,K lower.

We find a mean dust mass for these galaxies of $M_{\rm dust} \sim 3 \times 10^{8}$~$\rm M_{\sun}$, which is 
comparable to the characteristic mass in the dust mass function of $M^*_{\rm dust} \approx 4 \times 
10^8$~$\rm M_{\sun}$ measured at $z \sim 2.5$ by \citet*[][note the value quoted
here is taken from Table~3 of \citealt{Dunne_2011}]{Dunne_2003}. However, the range in $M_{\rm dust}$ spans
more than an order of magnitude, and the individual values are highly uncertain. The median $M_{\rm
dust}/M_*$ ratio for these galaxies is $\sim 5 \times 10^{-3}$ and spans the range $4 \times 10^{-4} - 3
\times 10^{-2}$. This is similar to the range found by \citet{RowanRobinson_2008}, with a sample reaching to
$z \sim 2$ and using a different method to estimate $M_{\rm dust}$. Fixing the value of $\beta = 2.0$ in
the modified blackbody model (equation~\ref{eq_blackbody}) would increase the mean dust mass that we find
by $\approx 60$ per cent.

\subsubsection{Joint optical-IR SED fitting}
\label{s_CIGALE}
We tested the sensitivity of the results described above to the simple modified blackbody model used in the
SED fitting (Section~\ref{s_fitting}) by jointly fitting the full optical-IR SEDs ($BVizH$ from \textit{HST},
IRAC channels 1-4, MIPS 24\,$\micron$, plus the \textit{Herschel} photometry) using \texttt{CIGALE}
\citep{Noll_2009}, a code which fits the attenuated optical light from stars and dust emission associated with
star formation and AGN simultaneously. The available models for use within \texttt{CIGALE} differ from those
assumed for deriving the GNS stellar masses (see Section~\ref{s_sample}) and the SFRs estimated in this work
(see Section~\ref{s_fitting}). We used the \citet{Maraston_2005} stellar population models to fit the optical
part of the spectrum, the \citet{DaleHelou_2002} templates to fit the dust emission, and a
\citet{Kroupa_2001} IMF. Some example SED fits are shown in Fig.~\ref{f_CIGALE}.

We find that \texttt{CIGALE} gives stellar masses that span the range $10.0 < \log (M_*/{\rm M_{\sun}}) < 11.5$, with median 
$\log (M_*/{\rm M_{\sun}}) = 10.9$, confirming that these systems have
high stellar masses, as measured in the GNS using a different SED fitting code \citep{Conselice_2011}. A two 
sample Kolmogorov-Smirnov (KS) test reveals that the stellar mass distributions are not significantly different ($p=0.33$), 
although there is a scatter of 0.23 dex in the residuals between the two stellar mass estimates for 
each galaxy.

The SFRs estimated by \texttt{CIGALE} are systematically lower than the results obtained using the modified blackbody
model, presumably as a result of the different stellar population model, IMF, and dust emission spectral templates used,
but the mean SFR ($210 \pm 30$~$\rm M_{\sun}$~yr$^{-1}$) is similar to that found from the modified
blackbody SED fits, despite this. The fraction of the IR luminosity due to
warm dust associated with AGN estimated by \texttt{CIGALE} spans the range $3-30$ per cent, with median
$\approx 5$ per cent. This suggests that star formation is the primary source of the IR emission in these
objects, as found in other studies \citep[e.g.][]{Netzer_2007, Mullaney_2011}.

\section{Stacking}
\label{s_stacking}
As shown in Section~\ref{s_crossMatching}, only 2.5 per cent of the $1.5 < z < 3$, $\log M_* (\rm M_{\sun}) >
9.5$ galaxy sample is detected in the 250\,$\micron$ maps used in this work, and the detected galaxies are
ULIRGs with large stellar masses ($\sim 10^{11}$~$\rm M_{\sun}$). We therefore performed a stacking analysis
to extend our study to galaxies with lower stellar masses and fainter far-IR luminosities. An additional 
advantage of the stacking analysis is that the results are less biased than those obtained from a small 
number of sources detected at low SNR. The stacking was performed on maps from which sources were not 
subtracted. Note that in contrast to the analysis in Section~\ref{s_detections}, additional maps at longer 
wavelengths than SPIRE were used in the stacking analysis (see Section~\ref{s_IRData}).

\begin{figure}
\includegraphics[width=8.5cm]{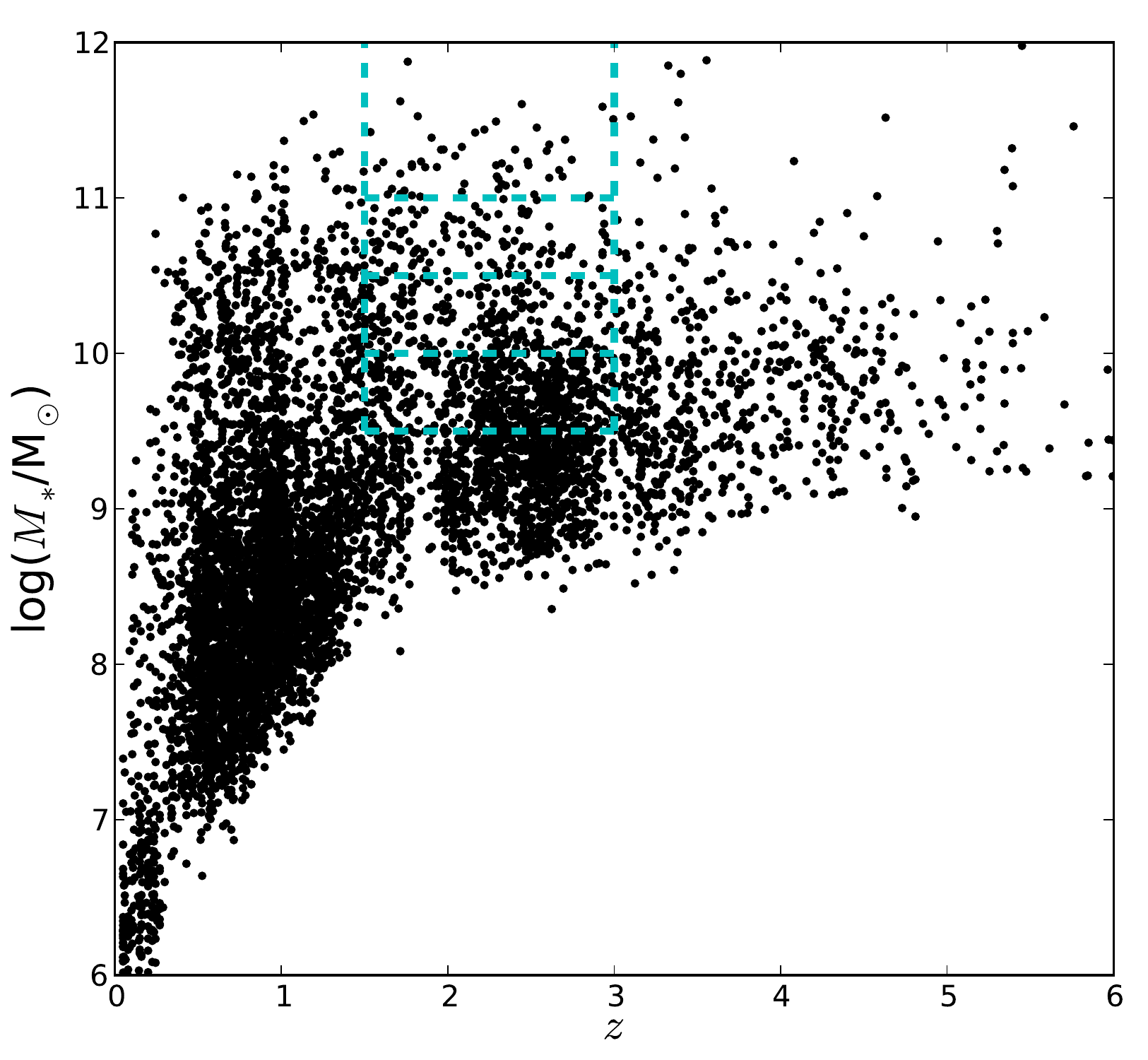}
\caption{Distribution of stellar masses with redshift for the GNS catalogue in both GOODS fields.
The blue dashed lines indicate the samples used in the stacking analysis presented in this paper.}
\label{f_samples}
\end{figure}

\begin{table*}
\caption{Properties of the mass limited galaxy samples for GOODS-North, GOODS-South, and the combined sample.
$N$ indicates the total number of galaxies that were stacked in each sample; $N_{z_{\rm spec}}$ is the number of these objects with
spectroscopic redshifts; $\langle z \rangle$ is the median redshift of the sample; $N_{\rm X}$ is the number of objects 
which are detected in X-rays (these are not included in the stacks and are not counted in $N$).}
\label{t_sampleProperties}
\begin{tabular}{|c|c|c|c|c|c|c|c|c|c|c|c|c|}
\hline
            & \multicolumn{4}{|c|}{North} & \multicolumn{4}{|c|}{South} & \multicolumn{4}{|c|}{Combined}\\
Mass Sample &   $N$ & $N_{z_{\rm spec}}$ & $\langle z \rangle$ & $N_{\rm X}$ & $N$ & $N_{z_{\rm spec}}$ & $\langle z \rangle$ & $N_{\rm X}$ & $N$ & $N_{z_{\rm spec}}$ & $\langle z \rangle$ & $N_{\rm X}$     \\
\hline
\phantom{0}$9.5 < \log (M_* /{\rm M_{\sun}}) < 10.0$        & 275   & 30    &2.4    & 0 & 233   & 20    &2.4    & 5 & 508   & 50    &2.4    & 5\\
$10.0 < \log (M_*/{\rm M_{\sun}}) < 10.5$      & 105   & 23    &2.2    & 5 & 111   & 17    &2.3    & 6 & 216   & 40    &2.3    & 11\\
$10.5 < \log (M_*/{\rm M_{\sun}}) < 11.0$      & 47    & 7 &2.2    & 10    & 53    & 11    &2.0    & 9 & 100   & 18    &2.1    & 19\\
\phantom{00011}$\log (M_* /{\rm M_{\sun}}) \geq 11$      & 25    & 0 &2.2    & 5 & 21    & 3 &2.1    & 4 & 46    & 3 &2.2    & 9\\
\hline
\end{tabular}
\end{table*}

\subsection{Sample definitions}
We divide the $1.5 < z < 3$ GNS galaxy sample into four bins of stellar mass, reaching to the $\log (M_*/{\rm M_{\sun}}) > 9.5$ 
limit to which the survey is complete \citep{Gruetzbauch_2011, Mortlock_2011}.
Fig.~\ref{f_samples} shows the location of the mass-limited subsamples in the ($M_*$,
$z$) plane, compared to the full GNS catalogue covering both GOODS fields. Due to the low SNR of the resulting
stacked detections (see Section~\ref{s_stackingResults}), we are not able to divide the sample into redshift
bins, nor examine subsamples of passive versus actively star forming galaxies (although note that the latter
is investigated using the GNS galaxy sample by \citealt{Bauer_2011}, using UV-based SFR measurements). 
Table~\ref{t_sampleProperties} lists the properties of the mass-limited subsamples we stack.

\subsection{Method}
\label{s_stackingMethod}
The far-IR data used in this work has low angular resolution, particularly in the SPIRE bands where the beam
sizes are 18$\arcsec$, 25$\arcsec$, and 36$\arcsec$ at 250, 350, and 500\,$\micron$, respectively, resulting in
relatively large confusion noise. The source densities of GNS galaxies per beam are also large (median 9
sources per beam at 250\,$\micron$), and if the effect of clustered confused sources is not accounted for, 
the resulting stacked fluxes will be biased. 

\begin{figure}
\includegraphics[width=8.5cm]{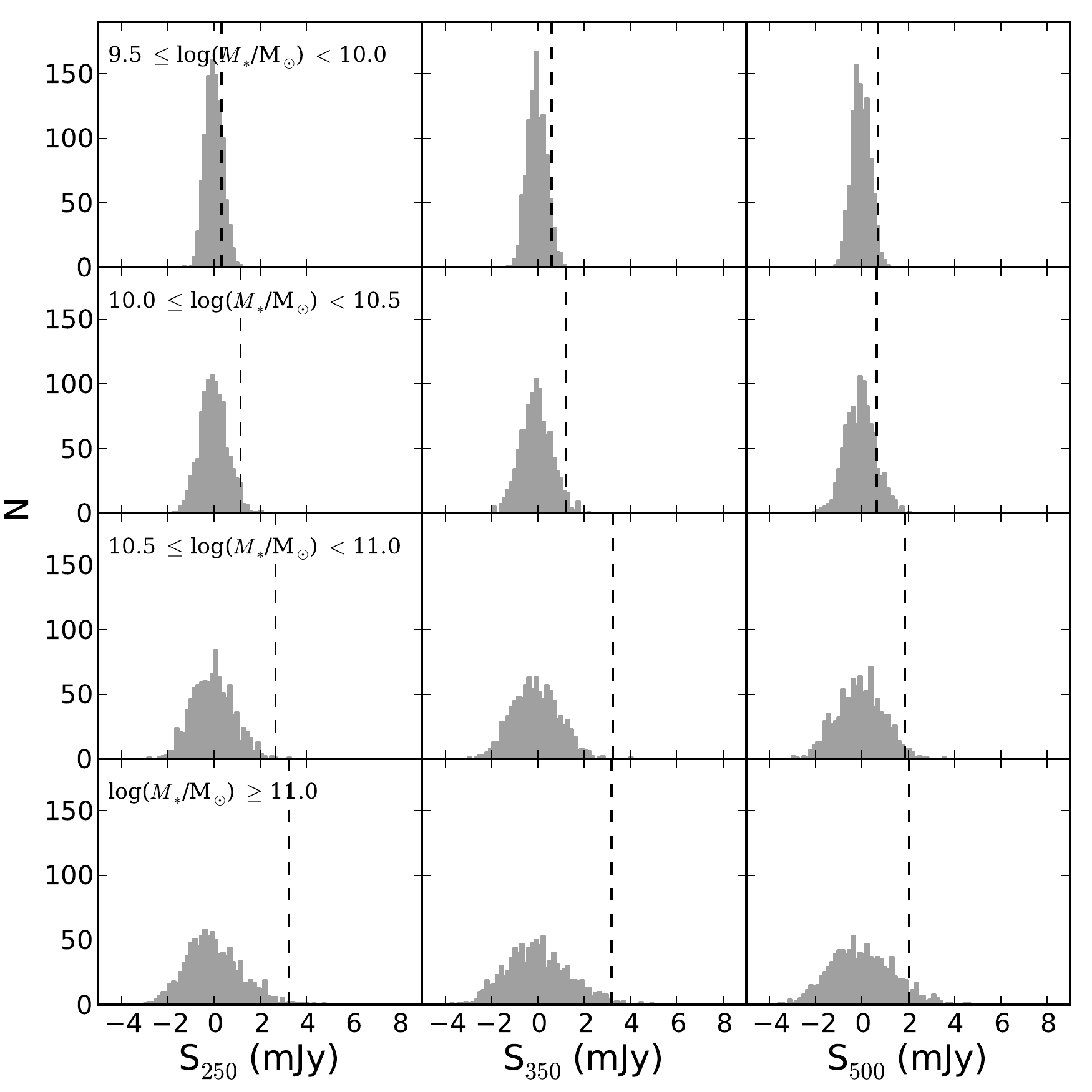}
\caption{Result of stacking on random positions for each stellar mass bin in the GOODS-N SPIRE maps. The
dashed line in each subplot indicates the stacked mean flux recovered when stacking on the real object
positions, as listed in Table~\ref{t_SEDResults}.}
\label{f_nullTest}
\end{figure}

\begin{figure*}
\includegraphics[width=8.5cm]{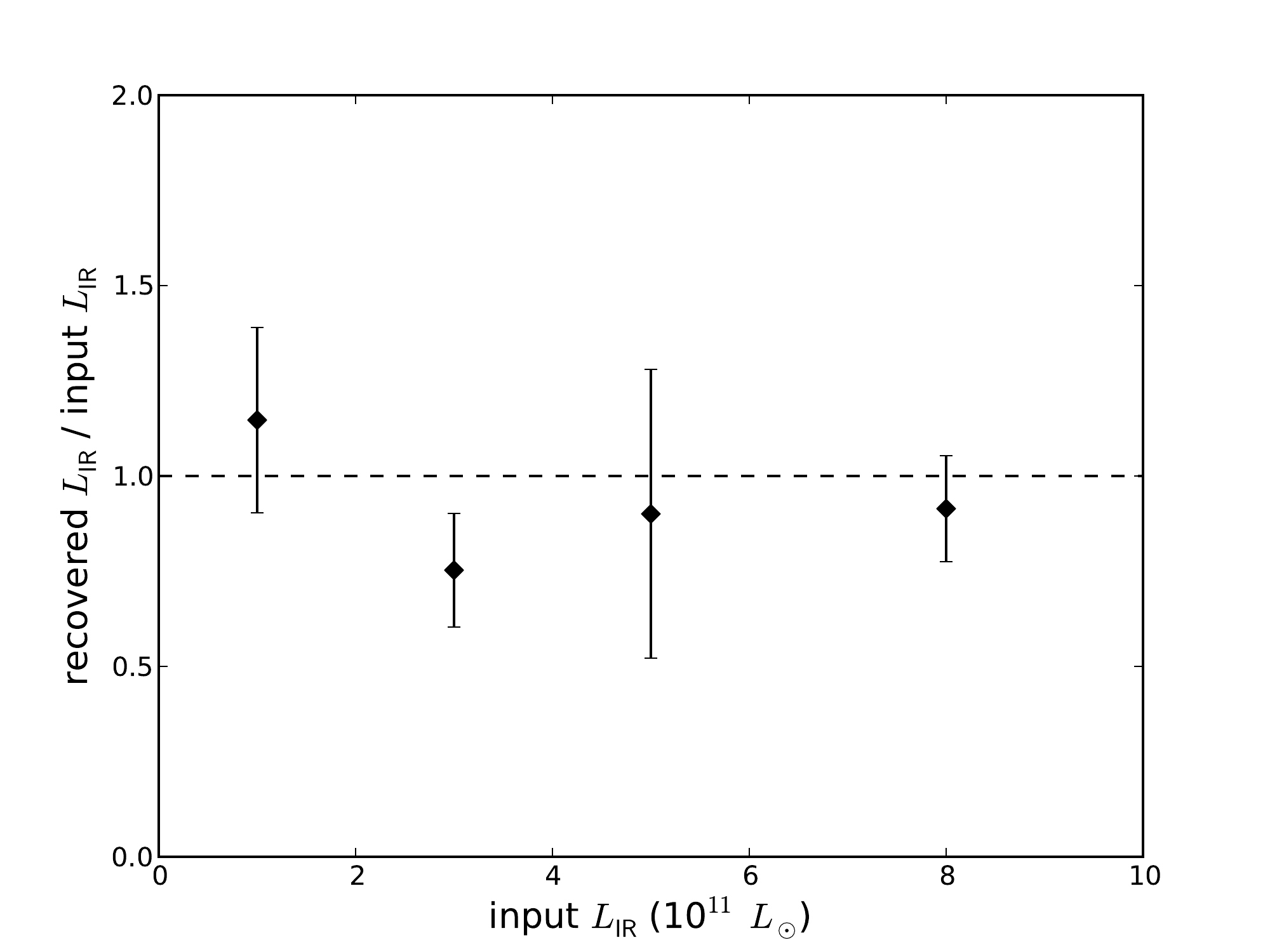}
\includegraphics[width=8.5cm]{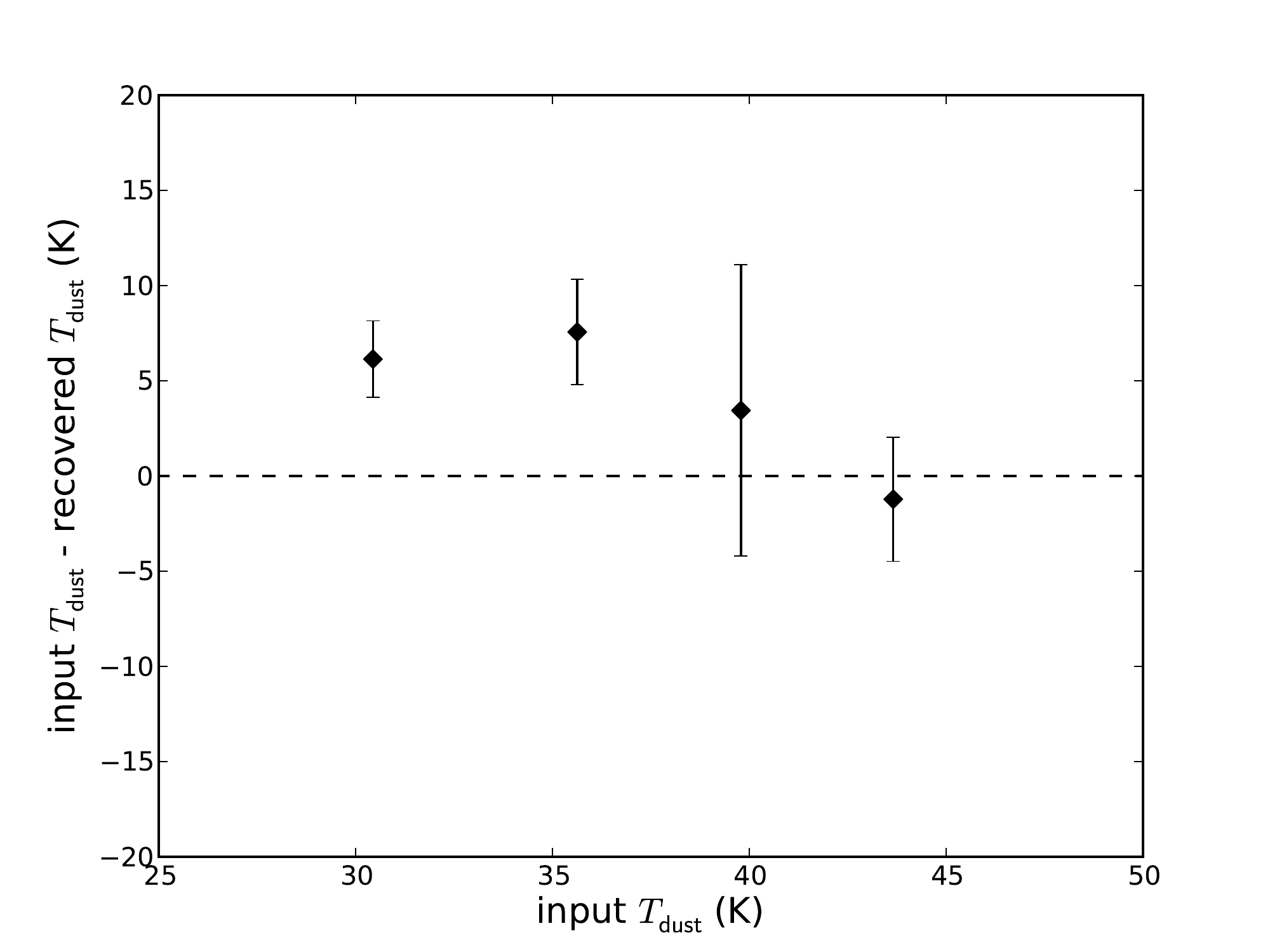}
\caption{Recovery of $L_{\rm IR}$ and $T_{\rm dust}$ when applying the stacking algorithm and SED fitting on
simple simulated maps.}
\label{f_sims}
\end{figure*}

\begin{figure}
\includegraphics[width=8.5cm]{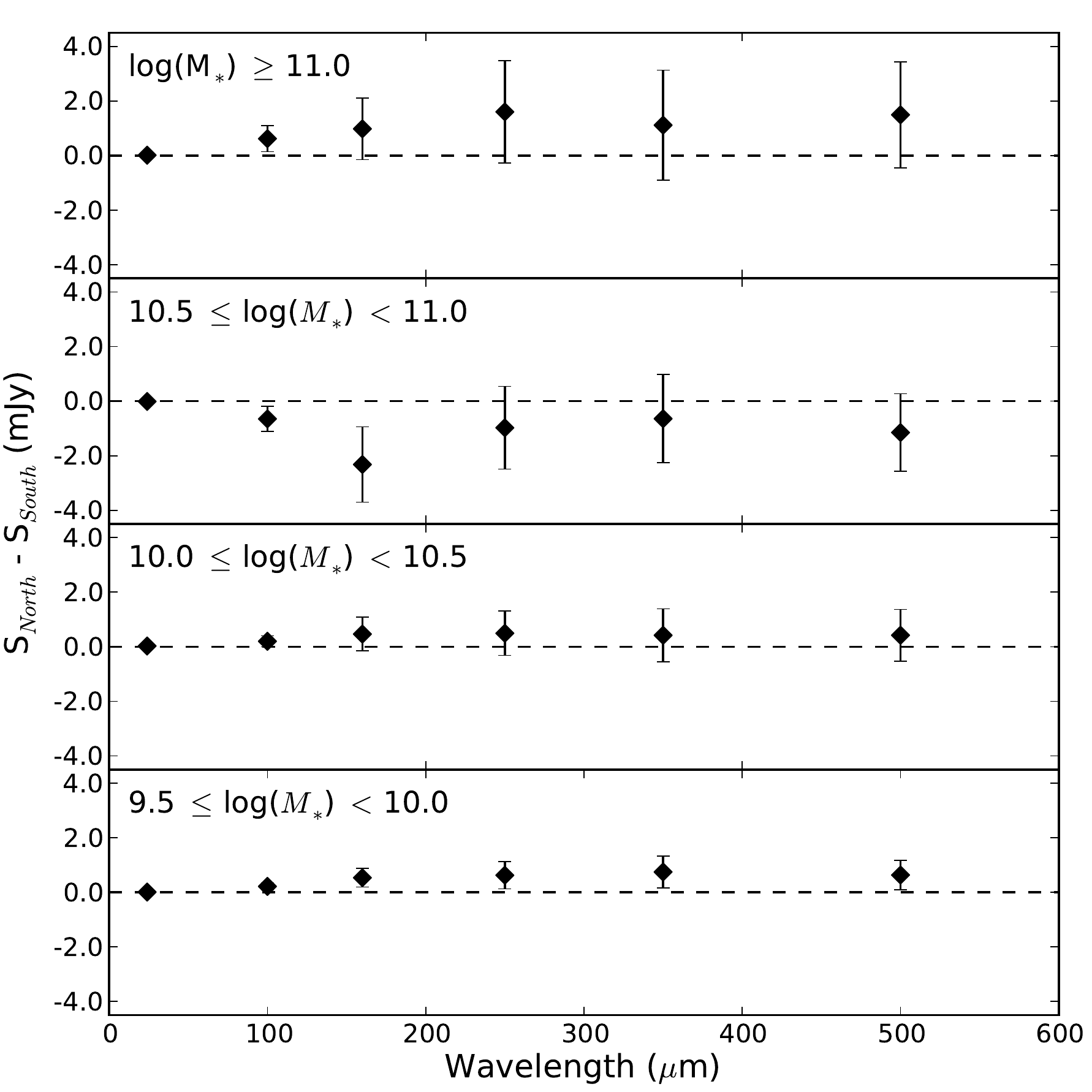}
\caption{Difference between the stacked flux densities in GOODS-N and GOODS-S for each stellar mass bin.
Within the large uncertainties there is no significant difference between the two fields, although the stacked
flux densities are generally fainter in GOODS-S.}
\label{f_N-SComparison}
\end{figure}

We use the global stacking and deblending algorithm of \citet[][KG2010 hereafter]{KurczynskiGawiser_2010} to
mitigate the effect of this bias \citep[for other approaches to this problem see][]{Bethermin_2012, Bourne_2012}. 
We generalised the method to simultaneously stack and deblend all of the
mass-limited samples (see Table~\ref{t_sampleProperties}), in addition to two `non-target' samples of objects.
The first of these non-target galaxy samples is drawn from the 24\,$\micron$ catalogue of \citet[][which is
also used to provide prior positions for source extraction in the \textit{Herschel} maps used in this
paper]{Magnelli_2009}. This catalogue provides coverage outside of the GNS footprint, and allows infrared
bright galaxies beyond the edges of the GNS fields to be deblended. Since 24\,$\micron$ bright sources are
correlated with sources detected at PACS and SPIRE wavelengths, these objects are the most likely to
contaminate stacked flux measurements of the mass-limited samples at far-IR wavelengths. The second non-target
galaxy sample consists of all GNS galaxies which are not 24\,$\micron$ sources and not
included in the stellar mass selected samples (i.e. with $z < 1.5$ or $z > 3$, and/or $\log (M_*/{\rm M_{\sun}}) <
9.5$). X-ray detected objects that are not included in the stellar mass selected samples
(Section~\ref{s_sample}) were also included in this second non-target sample.

We estimate errors on the stacked fluxes by bootstrapping: we run the stacking and deblending algorithm 1000
times, assigning the flux at each object position uniformly at random (with
replacement) from the observed fluxes in each sample. During this process, the
positions of all sources in the samples are kept fixed, and so the attenuation factors used in deblending
sources ($\alpha_{kj}$ in KG2010) remain constant (i.e. it is only the flux values that are bootstrap
resampled). We adopt the 68.3 percentile as the uncertainty in the stacked flux. We also estimated errors
by jackknifing (i.e. from the distributions of stacked fluxes obtained after
removing a single source from each stacking sample in turn), finding slightly smaller error bars -
the detection significances inferred using the jackknife error estimates are $0.1-0.2\sigma$ higher than those
obtained using the bootstrap error estimates.

We test the robustness of the mean stacked flux measurements by randomising the object positions
in each of the stacking samples (both target and non-target samples) and running the stacking algorithm,
repeating this process 1000 times. For simplicity, we perform this test using the GOODS-N sample only. We show
the results for each of the stellar mass samples in the SPIRE bands (since these are the most likely to suffer
from the effects of confusion as they have the largest beams) in Fig.~\ref{f_nullTest}. With the exception
of the lowest stellar mass bin, we find that the probability of a chance spurious stacked detection is higher
for the lower resolution channels. The detection probabilities inferred from this null test are consistent
with those obtained from stacking on real object positions and assuming the bootstrap error estimate; the
maximum difference is 0.3$\sigma$, with detection significances inferred from the random stack tests being
higher.

\subsubsection{Simulations}
\label{s_simulations}
We perform simple simulations to check that we can recover SED parameters such as $L_{\rm IR}$ and $T_{\rm dust}$ 
without significant bias. We create simulated maps with the same pixel scales as the real
GOODS-N maps and insert Gaussian sources with the appropriate FWHM for each channel at the positions of real
objects in the GNS catalogue. The simulated sources are modelled using the modified blackbody SED
(equation~\ref{eq_blackbody}). We note that this is somewhat idealised, as we do not include different SEDs
from those used in the fitting procedure.

For the stellar mass selected samples, anticipating the $L_{\rm IR}$ measurements obtained for the real maps (shown in
Section~\ref{s_stackingResults}), we set each model SED to have log~$L_{\rm IR} (\rm L_{\sun}) = $11.0, 11.5, 
11.7 and 11.9 for galaxies in stellar mass bins $\log (M_*/{\rm M_{\sun}})$ 9.5--10.0, 10.0--10.5, 10.5--11.0 and $> 11$,
respectively. We draw $T_{\rm dust}$ for each galaxy in each stellar mass subsample from a uniform distribution,
with a slightly different ($T^{\rm min}_{\rm dust}$--$T^{\rm max}_{\rm dust}$) range used for each bin: 
(15--45\,K), (20--50\,K), (25--55\,K), (30--60\,K), in ascending order of stellar mass. This ensures that 
each bin has different mean $T_{\rm dust}$, for clarity in the right panel of Fig.~\ref{f_sims}.

Models for galaxies in the non-target sample of 24\,$\micron$ bright sources have
log~$L_{\rm IR} (\rm L_{\sun}) = 11$, which is the median value we find for these sources when estimating their
$L_{\rm IR}$ from their 24\,$\micron$ flux densities alone (where we estimate $L_{\rm IR}$ for each source as
the median value over the full range of CE01 templates). We do not include the non-target galaxies that were not
detected at 24\,$\micron$ in the simulated maps. Each model source is redshifted to its corresponding $z$ in
the GNS catalogue. We apply a Gaussian random scatter of ($1+z_{\rm p}$)$\times 0.06$ in redshift to galaxies with
only photometric redshift estimates to simulate the effect of incorrect redshifts, where the amount of scatter
is as found by \citet{Gruetzbauch_2011} from a comparison of a subset of GNS galaxies with spectroscopic
redshifts (see Section~\ref{s_sample}). For sources in the 24\,$\micron$ detected non-target sample without
redshift information, we assign their model SED a redshift selected at random from the redshift distribution
of GNS galaxies detected at 24\,$\micron$. 

Fig.~\ref{f_sims} shows the results of running our stacking and SED fitting code
(Section~\ref{s_fitting}) on the simulated maps. We find that we recover $L_{\rm IR}$ to within
$\pm 30$ per cent down to the lowest stellar mass bin. We see that there is a small positive bias in 
$T_{\rm dust}$, with the recovered value being at most about 7\,K lower than the mean input $T_{\rm dust}$. 
This bias is absent if we set $T_{\rm dust}$ to a fixed value for all galaxies in each bin, and is likely to
be a consequence of the smearing of the stacked SED shape due to the different redshifts and dust temperatures
of the model SEDs that go into each stack.

\subsection{Results}
\label{s_stackingResults}

\begin{table*}
\caption{Stacked mean fluxes (in mJy) for $1.5 < z < 3$ GOODS NICMOS Survey galaxies in stellar mass bins. Ellipses 
(\nodata) indicate where the solution was negative and therefore unphysical.}
\label{t_SEDResults}
\begin{tabular}{|c|c|c|c|c|}
\hline
                    & \multicolumn{4}{|c|}{Sample: North}\\
Wavelength ($\mu$m) & $9.5 < {\rm log}(M_*) < 10.0$ & $10.0 < {\rm log}(M_*) < 10.5$ & $10.5 < {\rm log}(M_*) < 11.0$ & ${\rm log}(M_*) \geq 11.0$ \\
\hline
24          &   $0.008 \pm 0.005$&  $0.048 \pm 0.013$&  $0.056 \pm 0.011$&  $0.060 \pm 0.018$\\
100         &   $0.11 \pm 0.11$&    $0.26 \pm 0.20$&    $0.35 \pm 0.19$&    $0.84 \pm 0.43$\\
160         &   $0.22 \pm 0.26$&    $0.71 \pm 0.51$&    $0.53 \pm 0.59$&    $1.87 \pm 0.86$\\
250         &   $0.32 \pm 0.39$&    $1.15 \pm 0.64$&    $2.66 \pm 0.89$&    $3.23 \pm 1.47$\\
350         &   $0.59 \pm 0.47$&    $1.20 \pm 0.77$&    $3.23 \pm 1.10$&    $3.18 \pm 1.39$\\
500         &   $0.68 \pm 0.40$&    $0.64 \pm 0.76$&    $1.85 \pm 0.94$&    $2.03 \pm 1.38$\\
1160            &   $0.04 \pm 0.04$&    $0.03 \pm 0.07$&    $0.31 \pm 0.09$&    $0.41 \pm 0.17$\\
\hline
                    & \multicolumn{4}{|c|}{Sample: South}\\
Wavelength ($\mu$m) & $9.5 < {\rm log}(M_*) < 10.0$ & $10.0 < {\rm log}(M_*) < 10.5$ & $10.5 < {\rm log}(M_*) < 11.0$ & ${\rm log}(M_*) \geq 11.0$ \\
\hline
24          &   \nodata        &    $0.020 \pm 0.006$   &   $0.065 \pm 0.013$&  $0.049 \pm 0.025$\\
70          &   \nodata        &    $0.03 \pm 0.04$     &   $0.39 \pm 0.16$&    $0.04 \pm 0.10$\\
100         &   \nodata        &    $0.06 \pm 0.08$     &   $0.98 \pm 0.42$&    $0.22 \pm 0.22$\\
160         &   \nodata        &    $0.25 \pm 0.34$     &   $2.72 \pm 1.20$&    $0.92 \pm 0.70$\\
250         &   \nodata        &    $0.66 \pm 0.49$     &   $3.52 \pm 1.11$&    $1.67 \pm 1.21$\\
350         &   \nodata        &    $0.80 \pm 0.57$     &   $3.72 \pm 1.21$&    $2.15 \pm 1.58$\\
500         &   $0.06 \pm 0.36$&    $0.28 \pm 0.58$     &   $2.80 \pm 1.11$&    $0.62 \pm 1.36$\\
870         &   $0.04 \pm 0.07$&    \nodata             &   $0.29 \pm 0.19$&    $0.34 \pm 0.32$\\
\hline
                    & \multicolumn{4}{|c|}{Sample: Combined}\\
Wavelength ($\mu$m) & $9.5 < {\rm log}(M_*) < 10.0$ & $10.0 < {\rm log}(M_*) < 10.5$ & $10.5 < {\rm log}(M_*) < 11.0$ & ${\rm log}(M_*) \geq 11.0$ \\
\hline
24          &   $0.005 \pm 0.003$   &   $0.033 \pm 0.007$   &   $0.060 \pm 0.009$&  $0.056 \pm 0.016$\\
70          &   \nodata             &   $0.03 \pm 0.04$     &   $0.37 \pm 0.16$&    $0.07 \pm 0.09$\\
100         &   $0.01 \pm 0.06$     &   $0.16 \pm 0.10$     &   $0.66 \pm 0.23$&    $0.59 \pm 0.26$\\
160         &   \nodata             &   $0.48 \pm 0.31$     &   $1.60 \pm 0.68$&    $1.55 \pm 0.60$\\
250         &   $0.04 \pm 0.25$     &   $0.89 \pm 0.42$     &   $3.01 \pm 0.74$&    $2.68 \pm 0.95$\\
350         &   $0.26 \pm 0.29$     &   $0.95 \pm 0.48$     &   $3.38 \pm 0.82$&    $2.88 \pm 1.05$\\
500         &   $0.40 \pm 0.26$     &   $0.41 \pm 0.46$     &   $2.17 \pm 0.70$&    $1.59 \pm 0.95$\\
870         &   $0.04 \pm 0.07$     &   \nodata             &   $0.27 \pm 0.17$&    $0.31 \pm 0.27$\\
1160        &   $0.03 \pm 0.04$     &   $0.04 \pm 0.07$     &   $0.33 \pm 0.10$&    $0.43 \pm 0.19$\\
\hline
\end{tabular}
\end{table*}

Table~\ref{t_SEDResults} lists the mean stacked flux densities for each stellar mass selected subsample in 
each field. We find consistent results between the northern and southern fields given the large uncertainties,
although the stacked fluxes in the south are typically fainter than in the north for most stellar
mass samples (see Fig.~\ref{f_N-SComparison}). The stacked $S/N$ values are low: in the north, we obtain $\approx 2-3\sigma$
detections across almost all SPIRE and PACS bands for only the two most massive stellar mass bins. However,
the detection significance increases to $\approx 4\sigma$ in some channels for the second highest $\log M_*$
bin when the combined sample is used. The SNR in the lowest mass bin is only $\approx 1\sigma$ across the
PACS and SPIRE bands when using the combined sample.

\begin{figure}
\includegraphics[width=8.5cm]{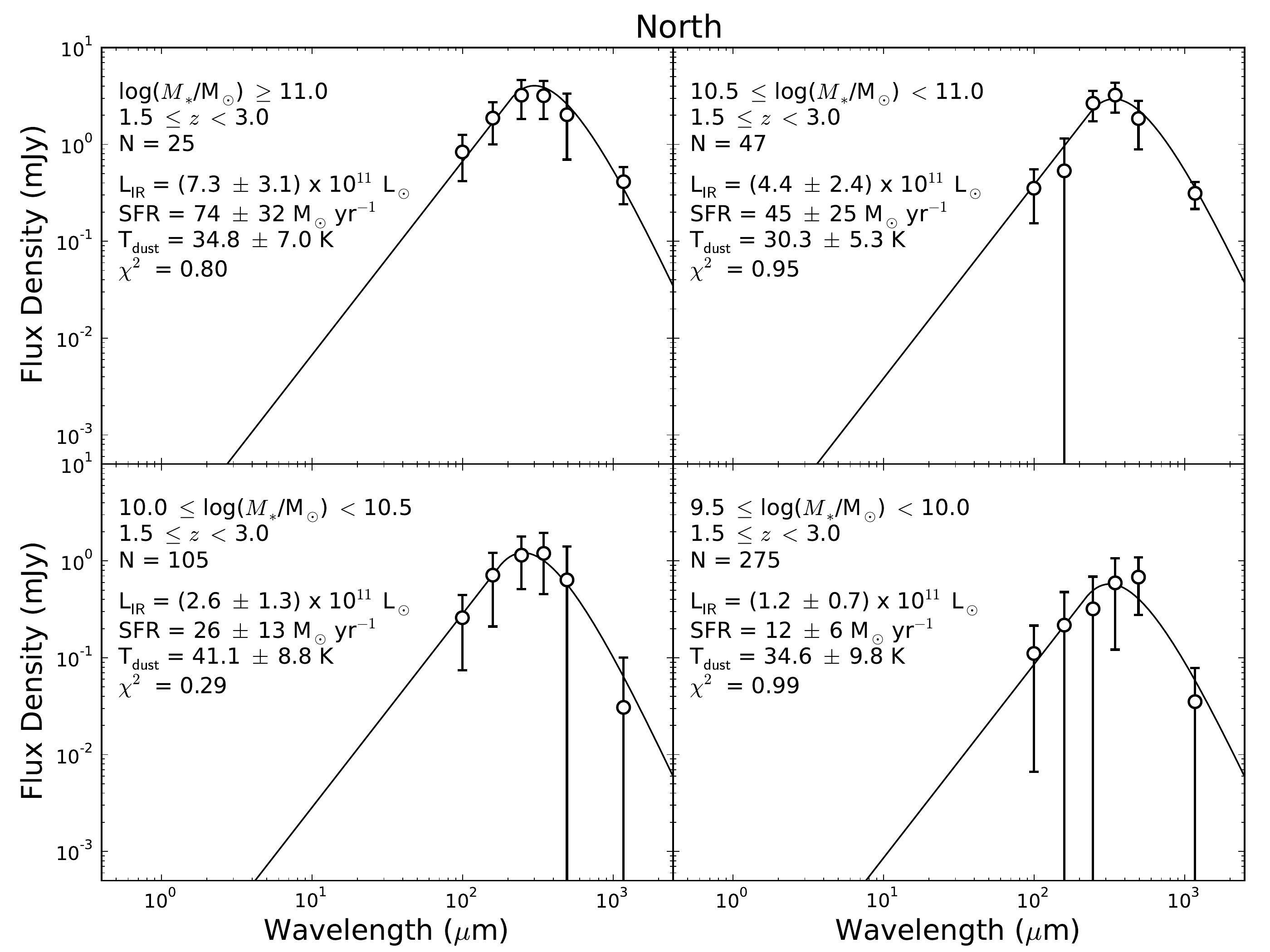}
\includegraphics[width=8.5cm]{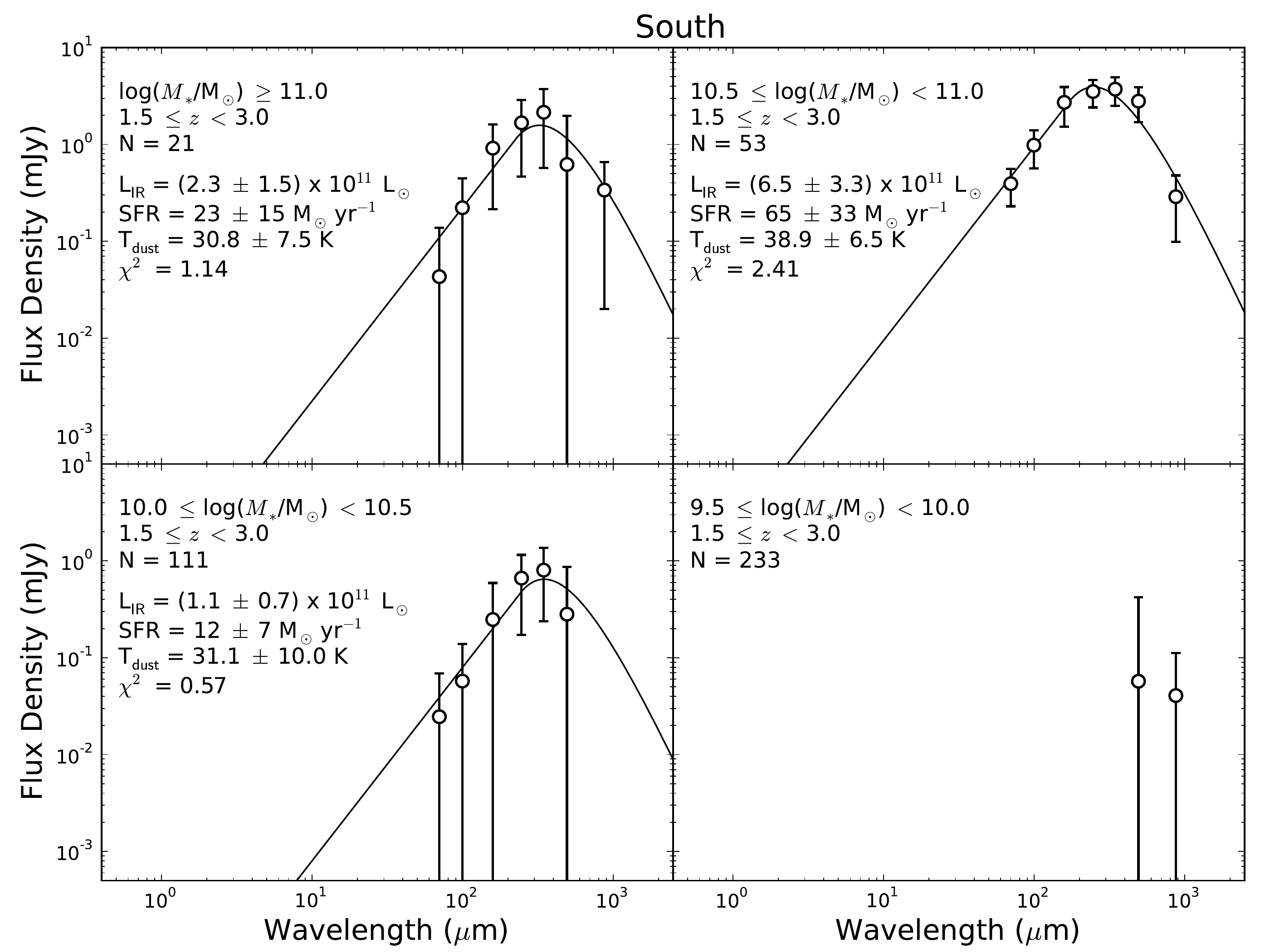}
\includegraphics[width=8.5cm]{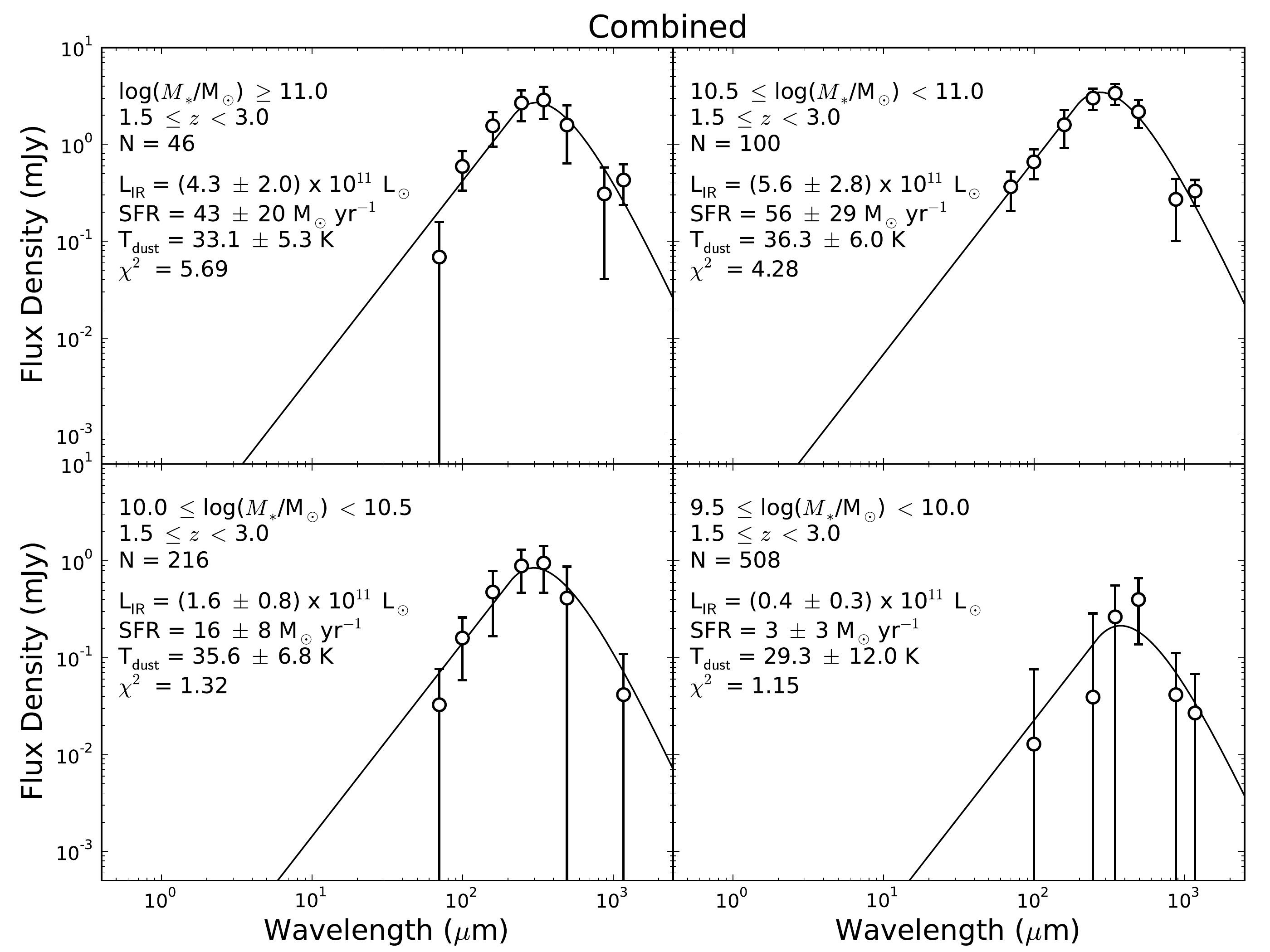}
\caption{The stacked far-IR/sub-mm SEDs as a function of stellar mass in GOODS-N (top), GOODS-S (middle) and
for both fields combined (bottom). Solid lines indicate the best-fitting modified blackbody model to each SED.}
\label{f_stackedSEDs}
\end{figure}

Despite the low SNR for each individual SED point, we proceed to fit the SEDs, in order to derive rough
estimates of $L_{\rm IR}$ and SFR$_{\rm IR}$ for each stellar mass bin. We include the low SNR points in the
fits, rather than excluding them, or treating them as upper limits. Under the assumption that the estimated error bars are 
reasonable (note that here they are obtained in a consistent way across all wavelengths), 
this should not bias the fit. We fit the SEDs for each stack using nearly the same method that was used for the SPIRE detected galaxies 
(Section~\ref{s_fitting}). We make one change to the fitting procedure in order to account for the wide redshift range covered by the
galaxy sample: during the Monte-Carlo procedure used to estimate error bars on the fitted parameters (i.e.
$L_{\rm IR}$, $T_{\rm dust}$, the uncertainties of which feed through to SFR$_{\rm IR}$ and $M_{\rm dust}$),
we bootstrap sample the redshift applied to the model SEDs from the distribution of redshift values in each stellar
mass bin. This approximately doubles the size of the uncertainties on $L_{\rm IR}$ and SFR in comparison to
those obtained when the redshift is held fixed at the mean redshift of the galaxy sample. 
Fig.~\ref{f_stackedSEDs} presents the stacked SEDs and best fit results using the modified blackbody 
templates for the northern, southern and combined samples. 

\subsubsection{Star formation}

We obtain estimates of SFR$_{\rm IR}$ for each sample with typically a factor of two uncertainty, despite the
low SNR measurements in each individual band. We find that
the difference between the stacked flux densities measured for the GOODS-N and GOODS-S fields
(Fig.~\ref{f_N-SComparison}) leads to lower SFRs for most stellar mass bins in the GOODS-S sample. However,
there is little tension between the SFRs measured in each field: the largest discrepancy is between the
highest stellar mass bins, but even in this case the difference in the SFR estimates is significant only at
the $<2 \sigma$ level. We find that the SFR$_{\rm IR}$ estimates obtained using the modified blackbody model
and the CE01 templates are consistent.

\begin{figure*}
\includegraphics[width=5.65cm]{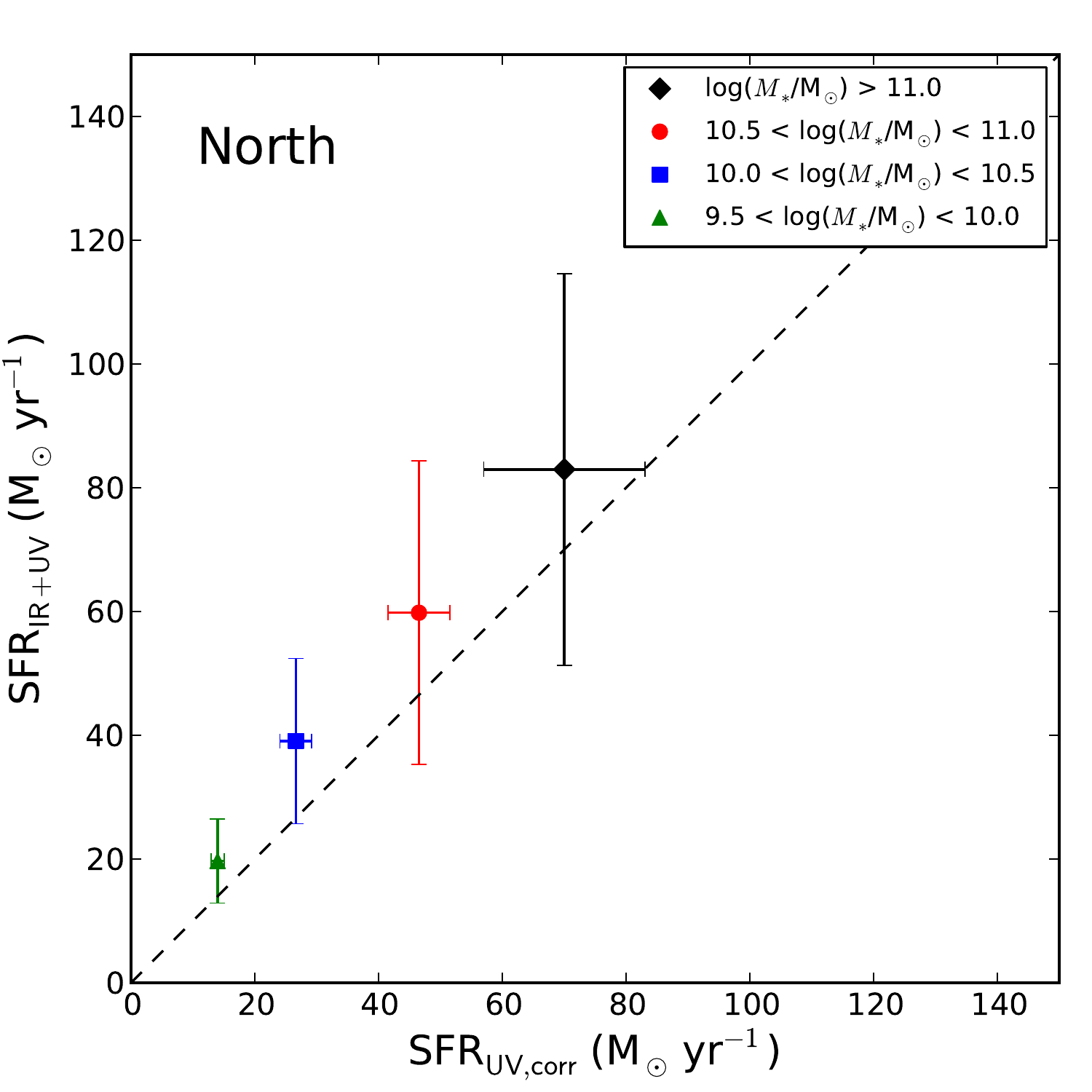}
\includegraphics[width=5.65cm]{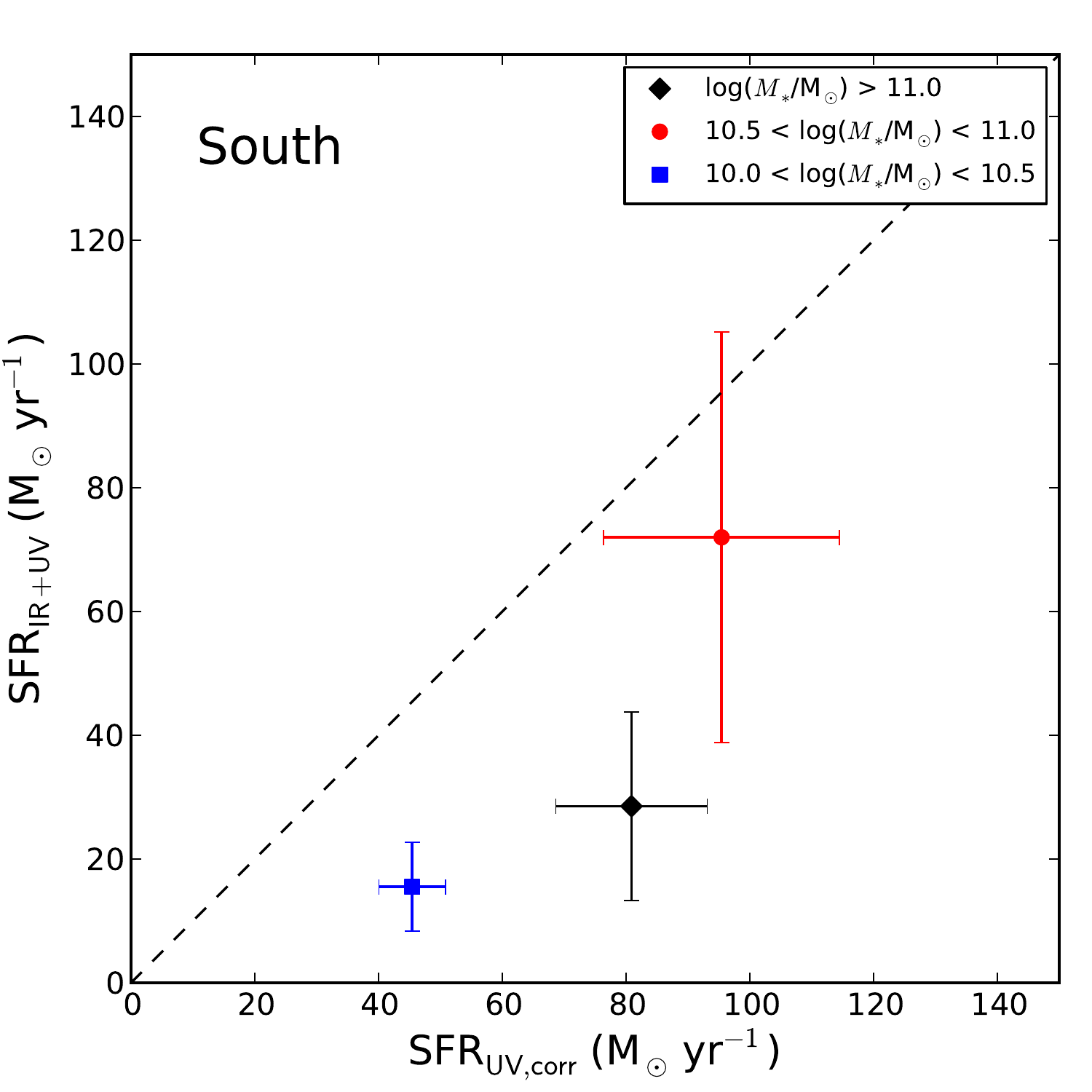}
\includegraphics[width=5.65cm]{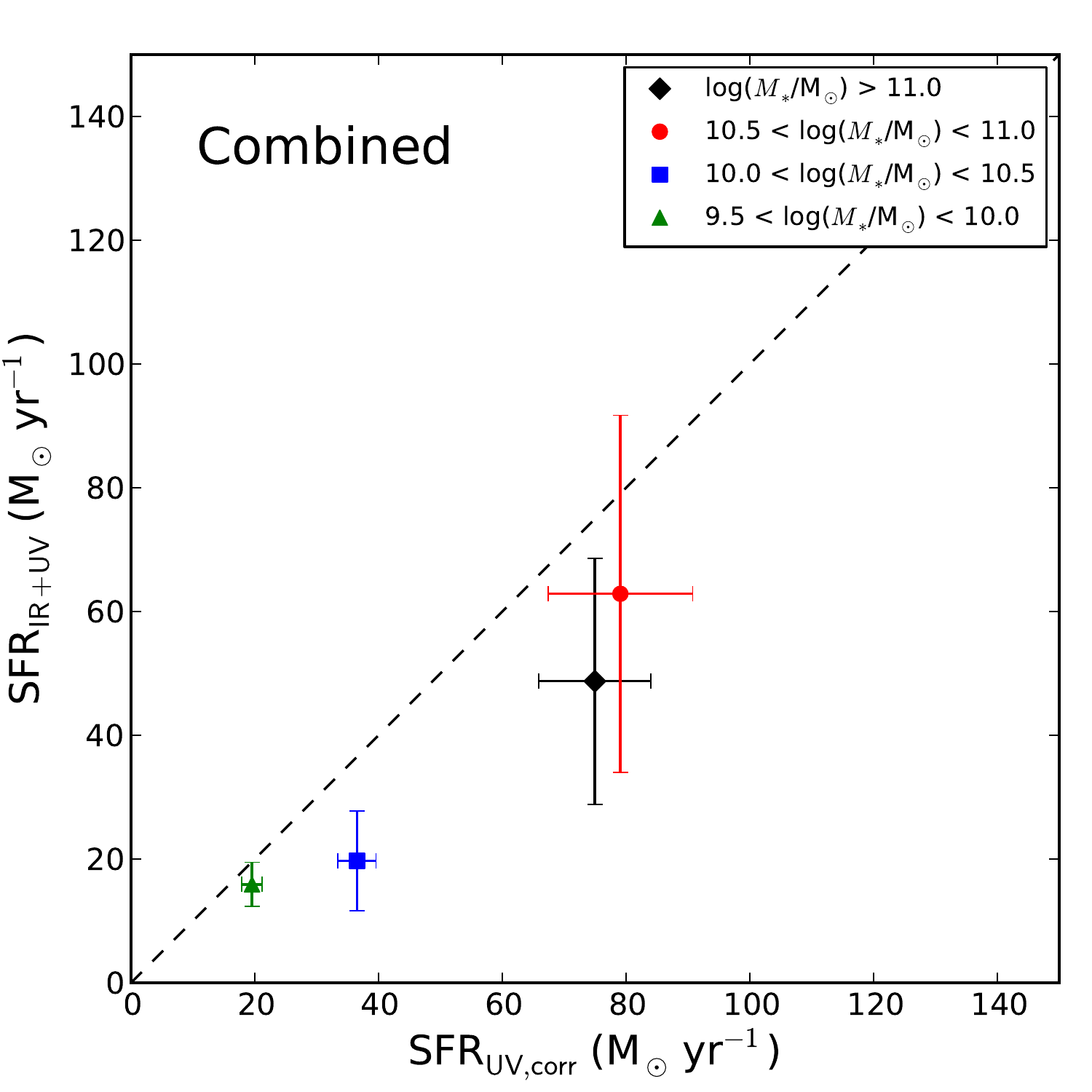}
\caption{Comparison of mean SFR in each stellar mass bin derived from stacking (SFR$_{\rm IR+UV}$; this work)
with the mean SFR derived from the UV-slope extinction corrected rest-frame UV flux (SFR$_{\rm UV,corr}$). The latter
uses measurements described in \citet{Bauer_2011}. We calculate the mean SFR$_{\rm UV,corr}$ using the same
galaxies as in the stellar mass bins used in the IR stacking analysis, after first scaling the
\citet{Bauer_2011} values to a \citet{Chabrier_2003} IMF. Results are shown for each GOODS field separately,
as well as the combined sample.}
\label{f_BauerSFRComparison}
\end{figure*}

We estimate mean total SFR$_{\rm IR+UV}$ for the stacked samples by adding to each sample the mean UV-based
estimate of unobscured SFR from \citet{Bauer_2011} for the same galaxies in each stellar mass bin.
Fig.~\ref{f_BauerSFRComparison} shows the resulting comparison with the mean UV-slope extinction corrected 
estimates (SFR$_{\rm UV,corr}$) from \citet{Bauer_2011} for the same galaxies. We see a rough agreement
between the two measurements given the large uncertainties, although while in GOODS-N SFR$_{\rm IR+UV}$ is
higher than SFR$_{\rm UV,corr}$, the opposite is true in GOODS-S. Much of this difference comes from a factor
$\sim 2$ difference in SFR$_{\rm UV,corr}$ between the two fields, with SFR$_{\rm UV,corr}$ being higher in
GOODS-S than GOODS-N. For all stellar mass bins apart from $\log (M_*/{\rm M_{\sun}}) > 11$, the difference in SFR$_{\rm
UV,corr}$ between the fields is significant at the $\approx 3\sigma$ level. The difference in SFR$_{\rm
IR+UV}$ between the fields is less significant, at most $1.6\sigma$. Also, in GOODS-S, the highest SFR
is seen for the 2nd most massive $\log\,M_*$ bin, in both SFR$_{\rm IR+UV}$ and SFR$_{\rm UV,corr}$,
although neither of these SFR estimates are significantly different from those measured for the most massive
$\log\,M^*$ bin.

We checked for differences between the GOODS-N and GOODS-S samples that could lead
to these effects. It is not likely that they arise from different redshift distributions: a two sample
KS test gives $p=0.21$, i.e. the distributions are not significantly different. Another
possibility is environmental effects: the GOODS-S field contains a galaxy overdensity at $z = 1.6$
\citep{Kurk_2009} which lies within our redshift range. This structure is thought to be a forming cluster of
galaxies, and so the denser environment on average relative to the GOODS-N field may lead to a higher fraction
of quiescent galaxies in GOODS-S, and therefore lower average SFR. However, we find that excising the region
within 2\,Mpc projected radius of this structure makes no significant difference to the derived SFRs. It
seems likely that the difference between the results for each field can be ascribed to the small area
covered by the GNS.

\subsubsection{SFR--$M_*$ relation}
\begin{figure*}
\includegraphics[width=5.65cm]{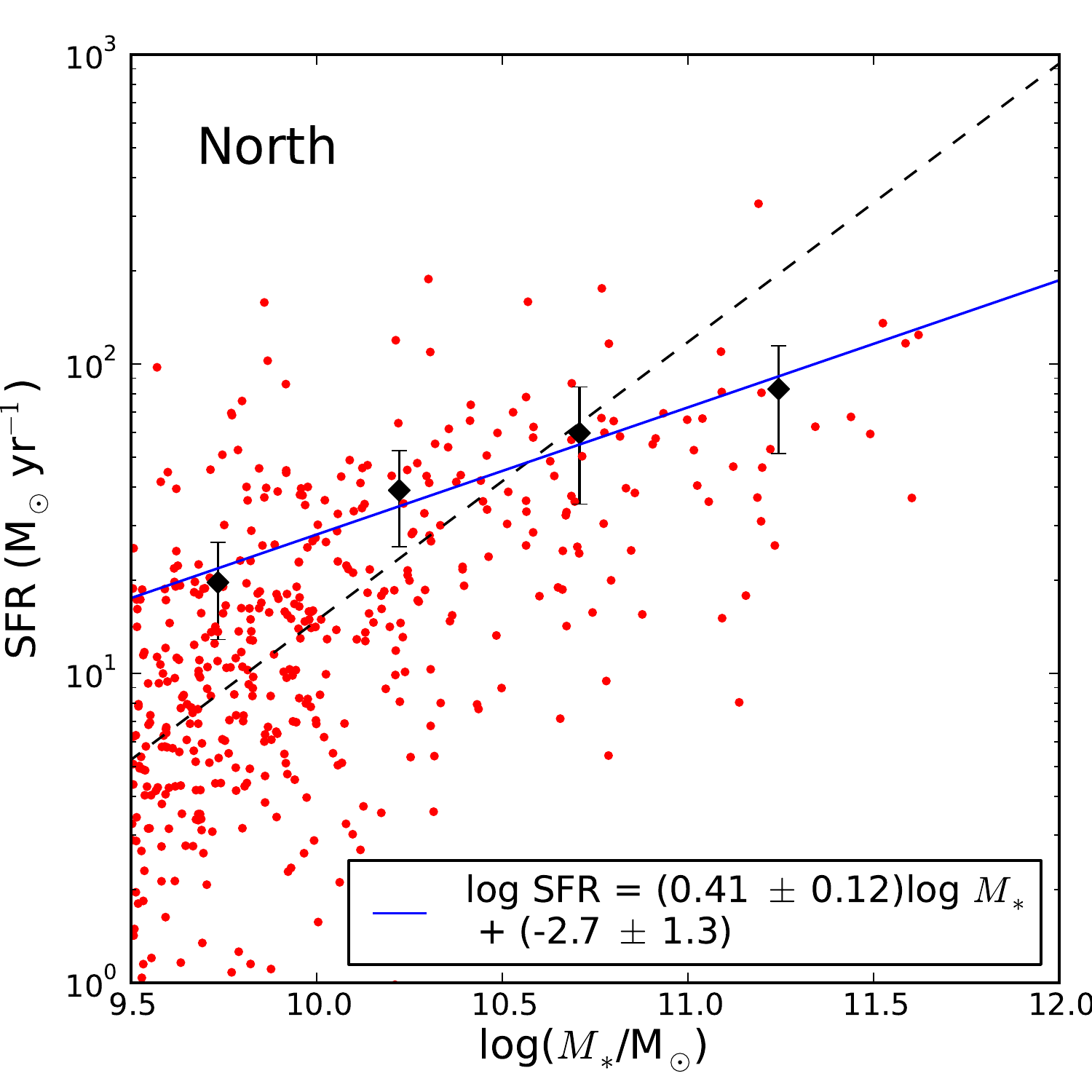}
\includegraphics[width=5.65cm]{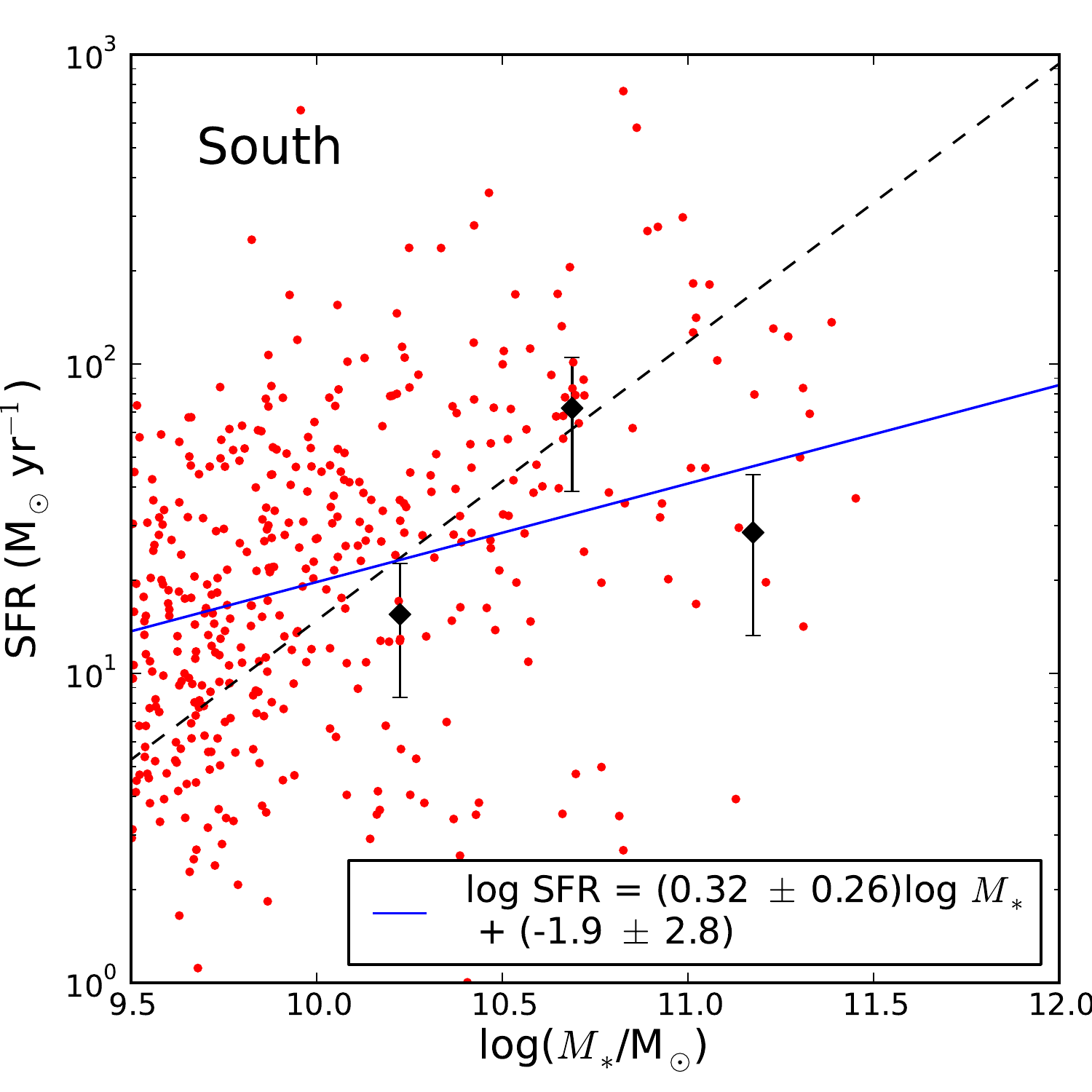}
\includegraphics[width=5.65cm]{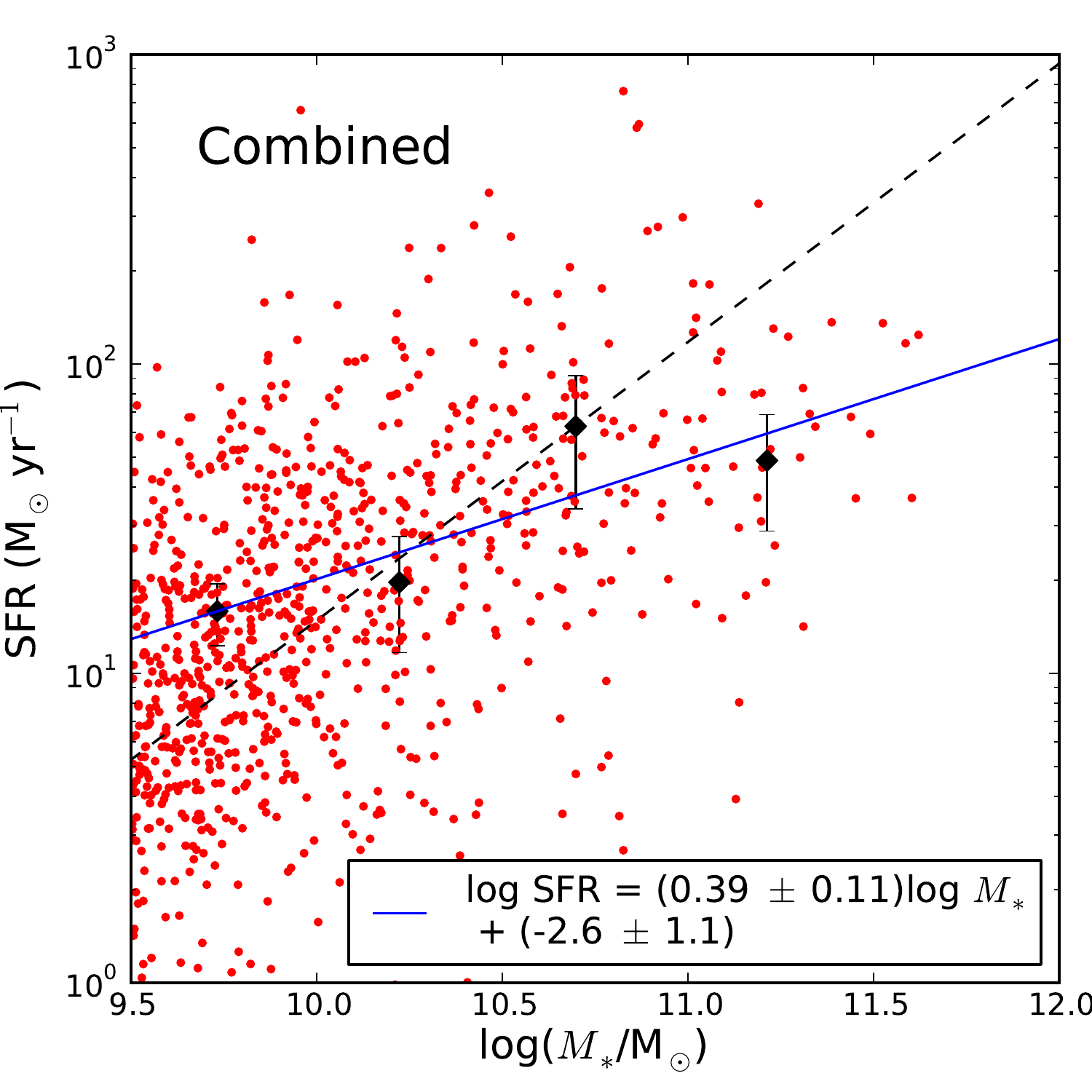}
\caption{The relation between SFR$_{\rm IR+UV}$ and $M_*$ for galaxies stacked in bins of stellar mass (black
diamonds). The blue line shows a weighted least squares fit to the relation.
The dashed line shows the SFR--$M_*$ relation measured by \citet{Daddi_2007I} at $z \sim 2$ for comparison. The
small red points show the SFR$_{\rm UV,corr}$ measurements for individual GNS galaxies from
\citet{Bauer_2011}, highlighting the large scatter in this relation. The results are shown for each GOODS
field separately, as well as the combined sample.}
\label{f_SFRMStarStacks}
\end{figure*}

Fig.~\ref{f_SFRMStarStacks} shows the SFR$_{\rm IR+UV}$-$M_*$ relation obtained for the stacks in each
field. Similarly to \citet{Bauer_2011}, we see a shallower relation compared to the
SFR--$M_*$ relation of \citet{Daddi_2007I}, who measured SFR$_{\rm UV, corr} \propto M_*^{0.9}$ for star
forming galaxies at $z \sim 2$. This is not surprising, because the GNS sample is selected differently,
purely by stellar mass, and therefore includes quiescent in addition to star forming galaxies 
\citep[see also][]{Bauer_2011}. Furthermore, the different methods used to measure SFR are also subject
to different selection effects. Using weighted
least squares regression, we find the relation:
\begin{align}
\log\,{\rm SFR_{UV+IR}} (\rm M_{\sun} \ {\rm yr^{-1}}) = (0.39 \pm 0.12) &\log\,(M_*/{\rm M_{\sun}})\nonumber  \\
&+(-2.5 \pm 1.2) \,
\end{align}
for the combined GOODS-N and GOODS-S fields. The fits obtained for the individual fields are indicated in
Fig.~\ref{f_SFRMStarStacks} and are consistent within the errors.

\begin{figure*}
\includegraphics[width=5.65cm]{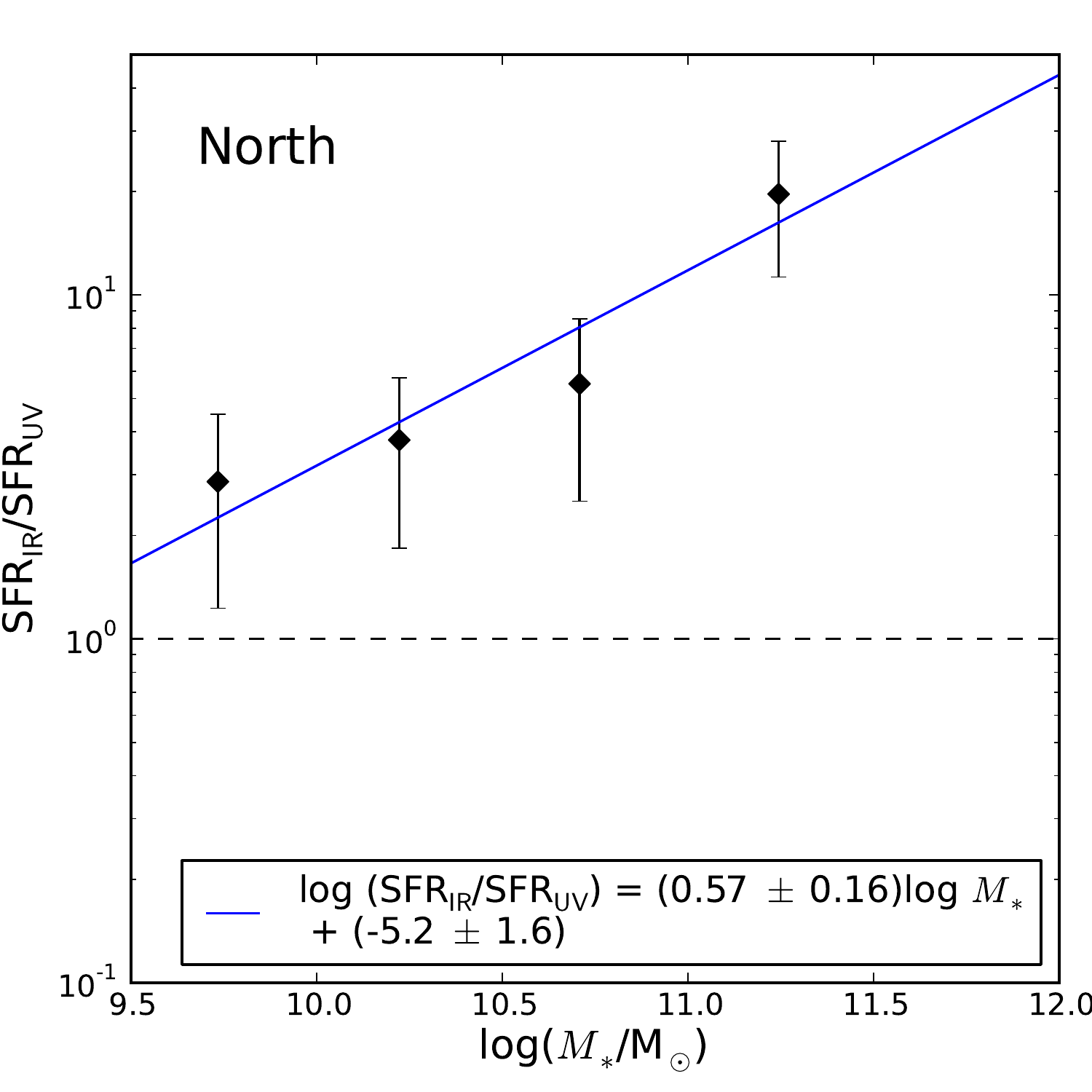}
\includegraphics[width=5.65cm]{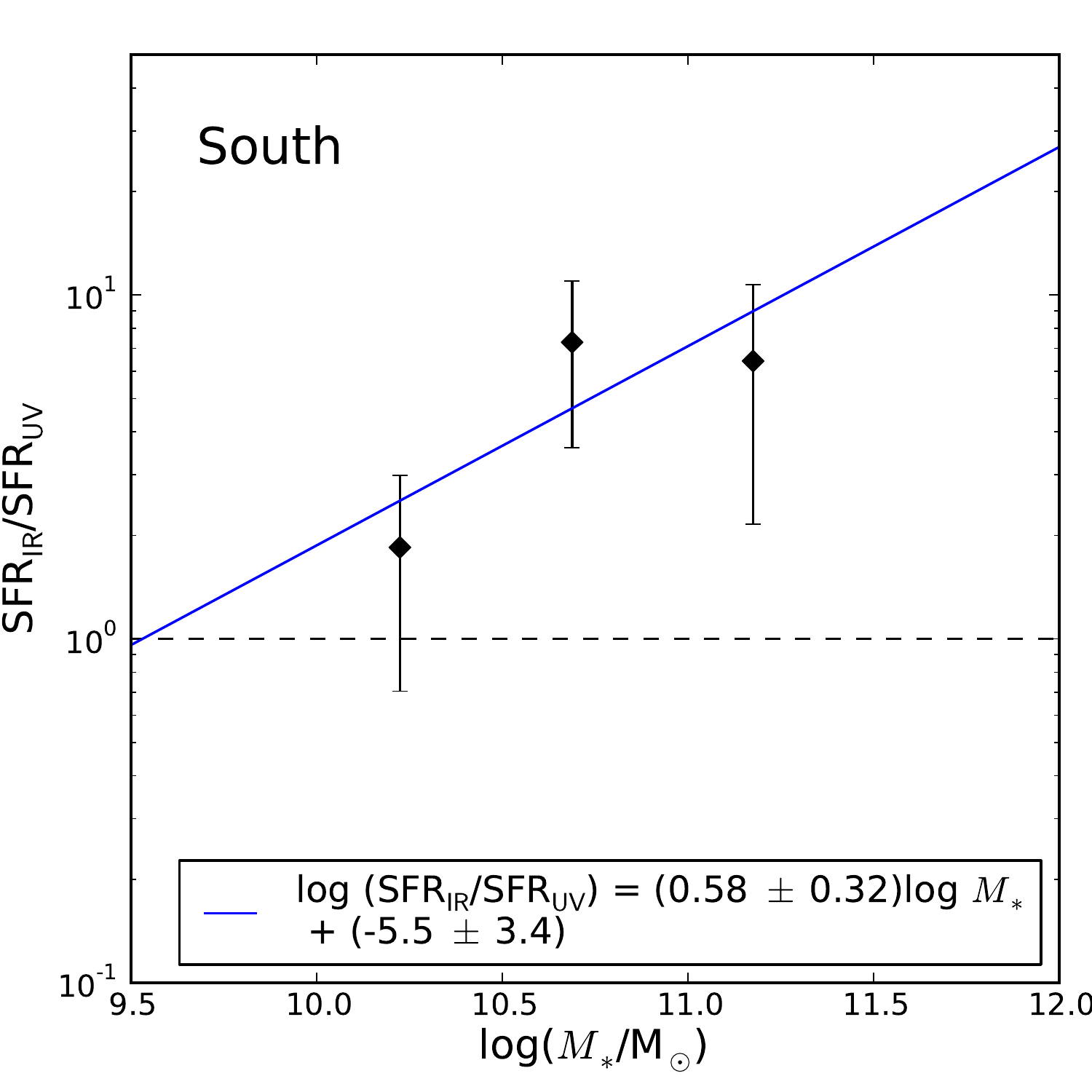}
\includegraphics[width=5.65cm]{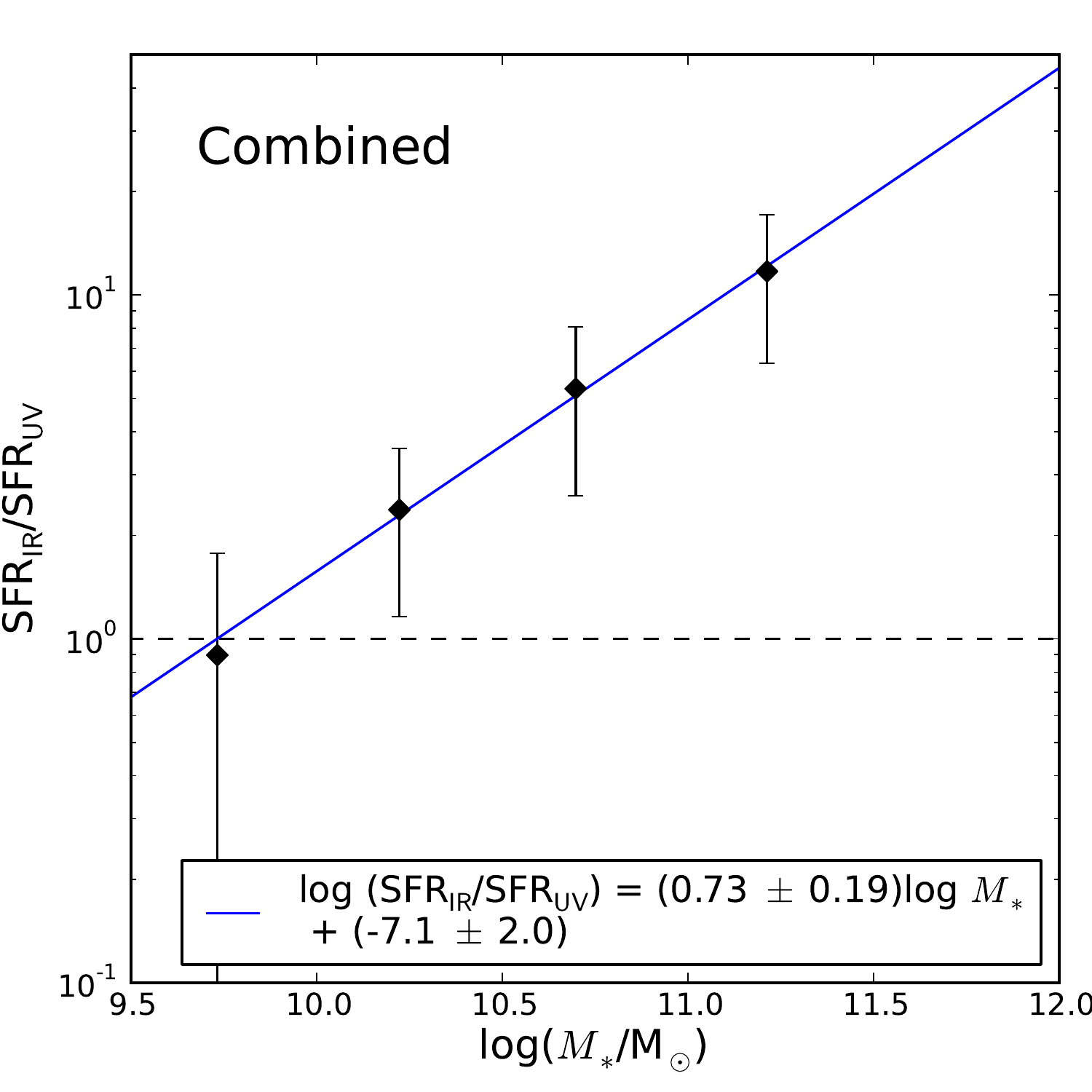}
\caption{The ratio of obscured to unobscured star formation (SFR$_{\rm IR}/$SFR$_{\rm UV}$) as a function of
$M_*$ for galaxies stacked in bins of stellar mass (where SFR$_{\rm UV}$ is taken from the measurements of
\citealt{Bauer_2011}). The blue line shows a weighted least squares fit to the
relation. The dashed line indicates SFR$_{\rm IR}/$SFR$_{\rm UV} = 1$. The results are shown for each GOODS
field separately, as well as the combined sample.}
\label{f_SFRRatioMStarStacks}
\end{figure*}

A straightforward comparison of our results with other works is not possible due to differences in sample
selection, stellar mass completeness and the wide redshift range used here. \citet{Karim_2011} performed a
stacking analysis in 1.4~GHz data using a 3.6\,$\micron$-selected sample of galaxies in the Cosmic Evolution
Survey field \citep[COSMOS;][]{Scoville_2007}; however, this survey suffers from incompleteness for 
$\log\,(M_*/{\rm M_{\sun}}) < 10.4$ at $z > 1.5$. Attempting a rough comparison of our measurements with this work, we find
that our SFR$_{\rm IR+UV}$ estimates are systematically lower, for similar stellar mass and redshift ranges.
However, the discrepancy is only significant at the 2--3$\sigma$
level for our most massive bin ($\log (M_*/{\rm M_{\sun}}) > 11$, where we find SFR a factor of $\sim 2$ less
than
\citealt{Karim_2011}), and there is reasonable agreement for the $10.5 < \log (M_*/{\rm M_{\sun}}) < 11$ bin.
\citet{Kurczynski_2010} studied a sample of star forming BzK galaxies (sBzKs) in the Extended
\textit{Chandra} Deep Field South, comparing several methods of measuring SFR using essentially the same stacking
algorithm we used in this work. Their sample was not stellar-mass-selected, but we find that our SFR$_{\rm
IR+UV}$ estimates for $\log (M_*/{\rm M_{\sun}}) > 10.5$ galaxies are in good agreement with their measurements
(obtained using IR data from MIPS, BLAST, and LESS) at the same redshift as our study, after accounting for
the \citet{Salpeter_1955} IMF assumed in \citet{Kurczynski_2010}. 

\subsubsection{Ratio of obscured to unobscured star formation and relation to stellar mass}

We plot the ratio of obscured to unobscured star formation (SFR$_{\rm IR}/$SFR$_{\rm UV}$) as a function of
stellar mass for the stacked samples in Fig.~\ref{f_SFRRatioMStarStacks}. Since the uncertainties are
large, this is not well constrained from our data. For both GOODS fields combined, we find the relation
\begin{align}
\log \,({\rm SFR_{IR} / SFR_{UV}}) = (0.69 \pm 0.19) &\log\,(M_*/{\rm M_{\sun}})\nonumber  \\
&+(-6.7 \pm 2.0) \,
\end{align}
using weighted least squares regression. As for the SFR--$M_*$ relation, the fits for the individual fields are
consistent within the large uncertainties. The slope of this relation suggests that galaxies with larger
stellar masses on average have a larger fraction of obscured star formation compared to lower mass galaxies. A similar
result is reported and discussed in \citet{Wuyts_2011b}, who suggest that the mass--metallicity relation is responsible, with
higher mass (metallicity) galaxies having larger dust column densities and correspondingly larger SFR$_{\rm IR}/$SFR$_{\rm UV}$
ratios \citep[see also][]{Pannella_2009}. For galaxies with $\log (M_*/{\rm M_{\sun}}) > 11$, 
we find the range spanned across the GOODS-N and GOODS-S fields is SFR$_{\rm IR}/$SFR$_{\rm UV} \sim 6-20$. For comparison, 
\citet{Reddy_2011} find SFR$_{\rm IR}/$SFR$_{\rm UV} = 4.2 \pm 0.6$ for a sample of $z \sim 2$ $L^*_{\rm UV}$ galaxies 
observed as part of the GOODS-\textit{Herschel} project.

\subsubsection{Dust properties}
Although we derive estimates for $T_{\rm dust}$ in each stellar mass bin from the SED fits
(Section~\ref{s_fitting}), they are not well constrained, with uncertainties $\sim $10\,K. All of the stellar
mass samples in each field have $T_{\rm dust}$ in the 20-40\,K range, consistent within errors across the
stellar mass range, and consistent with the mean value found for the individually detected sources
(Section~\ref{s_detectionsDust}). We note that simulations suggest that the $T_{\rm dust}$ estimates from
the stacked SEDs may be biased low, perhaps by roughly 7\,K (Section~\ref{s_simulations}). 

The estimates of $M_{\rm dust}$ we obtain are fairly low in comparison to $M^*_{\rm dust}$, the characteristic
mass in the dust mass function, as measured by \citet{Dunne_2011} for $0 < z < 0.5$ and at $z \sim 2.5$ by
\citet*{Dunne_2003}. The largest value of $M_{\rm dust}$ that we measure ($\approx 1.3 \times
10^8$~$\rm M_{\sun}$), corresponding to the $\log (M_*/{\rm M_{\sun}}) > 11$ bin, is a factor of $> 3$ lower than
$M^*_{\rm dust}$ measured by \citet{Dunne_2003} at similar $z$, and also lower than $M^*_{\rm dust}$ measured 
at $0.4 < z < 0.5$ \citep{Dunne_2011}. This may be as a result of the purely stellar mass based sample 
selection used here, which contains both passive and actively star forming galaxies; naturally, the samples
used in \citet{Dunne_2003, Dunne_2011} consist of galaxies selected in the sub-mm, and are therefore 
dominated by dusty star forming galaxies.

The relation we see between $M_{\rm dust}$ and $M_*$ is very poorly constrained ($\log~M_{\rm dust} \propto
\log~M_*^{0.45 \pm 0.37}$), owing to the large uncertainties in the dust masses, but suggests a mildly
decreasing $M_{\rm dust}/M_*$ ratio with increasing stellar mass, with $M_{\rm dust}/M_*$ falling from $\sim 
5 \times 10^{-3}$ to $\sim 7 \times 10^{-4}$ over the stellar mass range $9.5 < \log (M_*/{\rm M_{\sun}}) <
11$.

\section{Conclusions}
\label{s_conclusions}

We have investigated the far-IR properties of a stellar mass selected sample of $1.5 < z < 3$ galaxies
drawn from the GOODS NICMOS Survey - the deepest $H$-band \textit{HST} survey of its type prior to the 
installation of the WFC3 instrument - using deep \textit{Herschel} 70--500\,$\micron$ photometry from the HerMES and PEP key
projects. We found:

\begin{enumerate}

\item{Only 22 galaxies from the sample are detected at SNR $> 3$ at 250\,$\micron$. They are ULIRGs (median
log~$L_{\rm IR} (\rm L_{\sun}) = 12.4$), have high stellar masses (median $\log (M_*/{\rm M_{\sun}}) = 10.8$), and
are located at $z \approx 2$.
}
\vskip 3pt

\item{From fitting the SEDs of the SPIRE detected galaxies, we find they have mean SFR$_{\rm IR+UV}$ a factor of
$> 2$ higher than the UV-slope extinction corrected estimates of \citet{Bauer_2011}. However, we note that 
the IR-based SFR estimate suffers from a significant Malmquist bias, making the interpretation difficult. 
The mean dust temperature of the 16 objects with flux estimates in all HerMES and PEP bands 
($T_{\rm dust} = 35 \pm 6$\,K) is slightly lower than found for ULIRGs at $z < 1$.
}
\vskip 3pt

\item{Using a stacking algorithm which attempts to deblend sources, we find marginal detections ($2-4 \sigma$)
at SPIRE wavelengths when stacking the galaxy sample in bins of stellar mass, even for the highest
stellar mass bins ($\log (M_*/{\rm M_{\sun}}) > 10.5$).
}
\vskip 3pt

\item{Despite the low $S/N$ of the stacked flux measurements in each band, we obtain estimates of SFR$_{\rm
IR}$ for the stacked samples with factor $\sim 2$ uncertainties. We find that SFR$_{\rm IR+UV}$ measured for
the stacked samples is in reasonable agreement with measurements of SFR$_{\rm UV,corr}$ for the
same galaxy sample by \citet{Bauer_2011}. 
\vskip 3pt

\item{We find a relatively shallow slope for the SFR--$M_*$ relation (SFR$\ \propto M_*^{0.4 \pm 0.1}$)
compared to previous studies \citep[e.g.][]{Daddi_2007I}, which is likely due to selection effects, as our
purely stellar mass selected sample contains a mixture of passive and actively star forming galaxies.}
\vskip 3pt

\item{We find evidence for an increase in the ratio of obscured to unobscured star formation with
increasing stellar mass (SFR$_{\rm IR}/$SFR$_{\rm UV} \propto M_*^{0.7 \pm 0.2}$). This is most likely
a consequence of the mass--metallicity relation, with higher mass and metallicity galaxies being more
obscured.}
}
\vskip 3pt

Since the far-IR and sub-mm data used in this paper are amongst the deepest that will be obtained by 
\textit{Herschel}, it is clear that to make further progress in characterising the far-IR properties of low stellar mass
($\log (M_*/{\rm M_{\sun}}) < 10$) galaxies at $z \sim 2$ using \textit{Herschel}, a much larger galaxy sample is
needed, as will be provided by the Cosmic Assembly Near-infrared Deep Extragalactic Legacy Survey
\citep[CANDELS;][]{Grogin_2011, Koekemoer_2011}.
\end{enumerate}

\section*{Acknowledgments}

We thank the referee for many helpful comments which have improved this paper. We thank Amanda Bauer for 
providing the UV-based SFR measurements of GNS galaxies and useful discussions. MH and 
CJC acknowledge financial support from the Leverhulme Trust and STFC. Support for the GNS was also provided by
NASA/STScI grant HST-GO11082.

SPIRE has been developed by a consortium of institutes led by Cardiff Univ. (UK) and including: Univ. 
Lethbridge (Canada); NAOC (China); CEA, LAM (France); IFSI, Univ. Padua (Italy); IAC (Spain); Stockholm
Observatory (Sweden); Imperial College London, RAL, UCL-MSSL, UKATC, Univ. Sussex (UK); and Caltech, JPL, NHSC,
Univ. Colorado (USA). This development has been supported by national funding agencies: CSA (Canada); NAOC
(China); CEA, CNES, CNRS (France); ASI (Italy); MCINN (Spain); SNSB (Sweden); STFC, UKSA (UK); and NASA (USA).

PACS has been developed by a consortium of institutes led by MPE (Germany) and including: UVIE 
(Austria); KU Leuven, CSL, IMEC (Belgium); CEA, LAM (France); MPIA (Germany); INAF-IFSI/OAA/OAP/OAT, LENS,
SISSA (Italy); and IAC (Spain). This development has been supported by the funding agencies BMVIT (Austria),
ESA-PRODEX (Belgium), CEA/CNES (France), DLR (Germany), ASI/INAF (Italy) and CICYT/MCYT (Spain).

\bibliographystyle{mn2e}
\bibliography{refs}

\label{lastpage}

\end{document}